\documentclass[twocolumn,linenumbers]{aastex62}
\usepackage[utf8]{inputenc}
\usepackage{amsmath}
\usepackage{natbib}
\usepackage{graphicx}
\usepackage[normalem]{ulem}

\shorttitle{The Dark World}
\shortauthors {Fraine et al.}

\newcommand{\StarName}{WASP-43}
\newcommand{\PlanetName}{\StarName{}b}
\newcommand{\HST}{\textit{HST}}
\newcommand{\HSTSTIS}{\textit{HST}-STIS}
\newcommand{\HSTWFCTUVIS}{\textit{HST} WFC3/UVIS}
\newcommand{\HSTWFCTIR}{\textit{HST} WFC3/IR}

\newcommand{\Spitzer}{\textit{Spitzer}}
\newcommand{\JWST}{\textit{JWST}}
\newcommand{\Kepler}{\textit{Kepler}}

\newcommand{\fsed}{$f_{\rm sed}$}

\begin{document}
\nolinenumbers
\title{The Dark World: A Tale of \PlanetName{} in Reflected Light with \HSTWFCTUVIS{}}

\author[0000-0003-0910-5805]{Jonathan~Fraine}
\affiliation{Space Science Institute Center for Data Science}
\affiliation{Space Science Institute Center for Exoplanet and Planetary Science}
\author[0000-0002-4321-4581]{Laura~C.~Mayorga}
\affiliation{Center for Astrophysics \textbar{} Harvard \& Smithsonian, 60 Garden St, Cambridge, MA 02138, USA}
\affiliation{Johns Hopkins APL, 11100 Johns Hopkins Rd, Laurel, MD 20723, USA}
\author[0000-0002-7352-7941]{Kevin~B.~Stevenson}
\affiliation{Johns Hopkins APL, 11100 Johns Hopkins Rd, Laurel, MD 20723, USA}
\author[0000-0002-8507-1304]{Nikole~Lewis}
\affiliation{Department of Astronomy and Carl Sagan Institute, Cornell University, 122 Sciences Drive, Ithaca, NY 14853, USA}
\author[0000-0003-3759-9080]{Tiffany~Kataria}
\affiliation{California Institute of Technology - Jet Propulsion Laboratory}

\author[0000-0003-4733-6532]{Jacob~Bean}
\affiliation{University of Chicago Department of Astronomy and Astrophysics}
\author[0000-0002-3288-0802]{Giovanni~Bruno}
\affiliation{INAF - Catania Astrophysical Observatory, Via Santa Sofia, 78, I-95123 Catania, Italy}
\author[0000-0002-9843-4354]{Jonathan~J.~Fortney}
\affiliation{Department of Astronomy \& Astrophysics, University of California}
\affiliation{University of California, Santa Cruz, Other Worlds Laboratory}
\author[0000-0003-0514-1147]{Laura~Kreidberg}
\affiliation{Center for Astrophysics \textbar{} Harvard \& Smithsonian, 60 Garden St, Cambridge, MA 02138, USA}
\affiliation{Max Planck Institute for Astronomy, K\"onigstuhl 17, 69117, Heidelberg, Germany}
\author[0000-0002-4404-0456]{Caroline~V.~Morley}
\affiliation{University of Texas at Austin College of Natural Sciences}
\author[0000-0003-1609-5625]{Nelly~Mouawad}
\affiliation{Lebanese American University Department of Natural Science}
\author[0000-0002-9276-8118]{Kamen~O.~Todorov}
\affiliation{University of Amsterdam Anton Pannekoek Institute for Astronomy}
\author[0000-0001-9521-6258]{Vivien~Parmentier}
\affiliation{University of Oxford Department of Physics}
\author[0000-0003-4328-3867]{Hannah~.R.~Wakeford}
\affil{School of Physics, University of Bristol, HH Wills Physics Laboratory, Tyndall Avenue, Bristol BS8 1TL, UK}

\author[0000-0002-5032-5060]{Y.~Katherina~Feng}
\affiliation{Department of Astronomy \& Astrophysics, University of California}
\affiliation{NSF Graduate Research Fellow}
\affiliation{University of California, Santa Cruz, Other Worlds Laboratory}
\author[0000-0003-4220-600X]{Brian~M.~Kilpatrick}
\affiliation{Space Telescope Science Institute}
\author[0000-0002-2338-476X]{Michael~R.~Line}
\affiliation{Arizona State University, School of Earth and Space Exploration}

\begin{abstract}
    \nolinenumbers
    Optical, reflected light eclipse observations provide a direct probe of the exoplanet scattering properties, such as from aerosols.
    We present here the photometric, reflected light observations of \PlanetName{} using the \HSTWFCTUVIS{} instrument with the F350LP filter (346-822nm) encompassing the entire optical band.
    This is the first reflected light, photometric eclipse using UVIS in scanning mode; as such we further detail our scanning extraction and analysis pipeline \texttt{Arctor}.
    Our \HSTWFCTUVIS{} eclipse light curve for WASP-43 b derived a 3-$\sigma$ upper limit of 67~ppm on the eclipse depth, which implies that \PlanetName{} has a very dark dayside atmosphere.
    With our atmospheric modeling campaign, we compared our reflected light constraints with predictions from global circulation and cloud models, benchmarked with \HST{} and \Spitzer{} observations of \PlanetName{}.
    We infer that we do not detect clouds on the dayside within the pressure levels probed by \HSTWFCTUVIS{} with the F350LP filter (P~$>$~1~bar).
    This is consistent with the GCM predictions based on previous \PlanetName{} observations.
    Dayside emission spectroscopy results from \PlanetName{} with \HST{} and \Spitzer{} observations are likely to not be significantly affected by contributions from cloud particles.
\end{abstract}


\section{Introduction}
The climates of planets in the solar system and beyond are strongly shaped by the presence of aerosols (clouds and/or hazes) in their atmospheres. These clouds and hazes can reflect starlight back into space, absorb photons from across the stellar spectrum, and emit at longer wavelengths \citep{Marley1999ApJ, Sudarsky2000ApJ}. 
Because aerosols are expected to efficiently reflect starlight into space, aerosol properties can be constrained through planetary optical and near-ultraviolet (NUV) reflection spectra, often 
called albedo spectra. Optical/NUV observations of Solar System objects are abundant; but only a handful of exoplanets have been measured in reflected light \citep{Evans2013ApJ,Heng2013ApJ,Angerhausen2015PASP, Esteves2015ApJ,Shporer2015AJ, Haggard2018MNRAS}.

Transiting exoplanet eclipse observations provide a method for direct detection of photons from an exoplanet's atmosphere.
Eclipse measurements derive the dayside brightness of the exoplanetary atmosphere, which in the optical and NUV 
provides a measure of the geometric albedo of the planet. From such measurements, we can infer the compositions of the materials (clouds, hazes, molecules, and atoms) that most strongly shape the albedo spectrum of the planet \citep{Marley1999ApJ}.

To date, the presence of aerosols in exoplanetary atmospheres has largely been inferred through the muting of molecular spectroscopic features at infrared wavelength and scattering slopes, or lack thereof, at visible wavelengths seen in transmission spectra {(e.g.~\citealp{Esteves2015ApJ, Nikolov2015MNRAS, Nortmann2016AA, Nortmann2018Sci, Sing2016Nature, Palle2017AA, Wakeford2017MNRAS,  Allart2017AA, Allart2019AA, Chen2017AAa, Chen2017AAb, Chen2018AA, Chen2020AA, vonEssen2018AA, Spake2018Natur, Todorov2019AA, vonEssen2020AA}).}
{Because of its short (19.5 hour) orbital period, the hot-Jupiter WASP-43b \citep{Hellier2011AA} has been a prime target for phase-curve observations at infrared wavelengths with both Hubble \citep{Stevenson2014Science} and Spitzer Space Telescopes \citep{Stevenson2017AJ}. These publications, and related analyses \citep{Kataria2015ApJ}, invoked aerosols to explain discrepancies between predictions and observations for exoplanetary phase curve behaviour} {\citep{Stevenson2014Science, Kataria2015ApJ, Oreshenko2016MNRAS, Stevenson2017AJ, Haggard2018MNRAS}}.
Note that because the eclipse was detected in several IR observations, we can rule out the possibility that {a non-detection (see below), in our \HSTWFCTUVIS{} data, described here,} would be caused by eccentricity modifying the eclipse time.

These phase-curve observations found little flux to be emanating from the night-side of the planet, suggesting thick cloud coverage. 
Furthermore, \cite{Keating2017ApJ} showed that reflected light eclipses can act as a pole-arm for interpreting models from infrared observations, such as with the existing (i.e.~\HST{} \& \Spitzer{}) and future {\it James Webb Space Telescope} (\JWST{}) observations. 

\cite{Kataria2015ApJ} explored the potential for cloud formation to influence the phase-curve observations of hot-Jupiter planets like \PlanetName{}, which found that clouds could be a significant source of contamination in these observations.
In canonical exoplanet cloud formation theory, using thermochemical equilibrium, planets with $T_{eq} < 1650K$ are predicted to form substantial cloud layers of 
silicate materials, such as enstatite (MgSiO$_3$) or forsterite (Mg$_2$SiO$_4$) \citep{Ackerman2001, Marley2013Book}.
The optical reflectivity of these silicates can be large enough that, under the most ideal conditions, observational simulations have shown that they could produce a geometric albedo $>$ 0.5.
Furthermore, a geometric albedo of 0.5 for this hot Jupiter would produce NUV-optical eclipse depths upwards of 500~ppm {\citep{Oreshenko2016MNRAS, Parmentier2016ApJ}}. 
In contrast, the majority of hot Jupiter models agree that albedos will be very low without clouds, providing an implicit metric for a priori cloud existence.


Here we present the first optical to NUV eclipse observations of \PlanetName{}.
These observations leverage a mode on \HST{} that has not previously been used
in the study of exoplanet eclipse observations. 
We selected \HST{} for this study because of its photometric (RMS$_{\text{reduced}}$=172~ppm) and pointing ($\Delta X \sim \Delta Y<$0.05 pixels) stability, as well as the capacity for our achieved eclipse depth precision ($\delta_{\text{eclipse depth}}\sim$34~ppm).

In the following sections (\S2), we present the operations of \HST{}'s scanning mode observations. In \S3 we discuss our newly developed \texttt{Arctor} pipeline for trace photometry (i.e.~scanning mode arcs or asteroid streaks; see \autoref{appendix:arctor} as well). 
We present our results in \S4, including comparisons to other known, high precision, reflected light, eclipse depth measurements. \S5 discusses our modeling campaign to consider how our upper limit provides contest to cloud particle formation models.
In \S6, we continue with a discussion about how to use further reflected light observations of hot Jupiters to constrain cloud formation physics \citep{Oreshenko2016MNRAS}.
And, finally, in \S7, we present our conclusions.

\begin{figure}
\centering
\includegraphics[width=\linewidth]{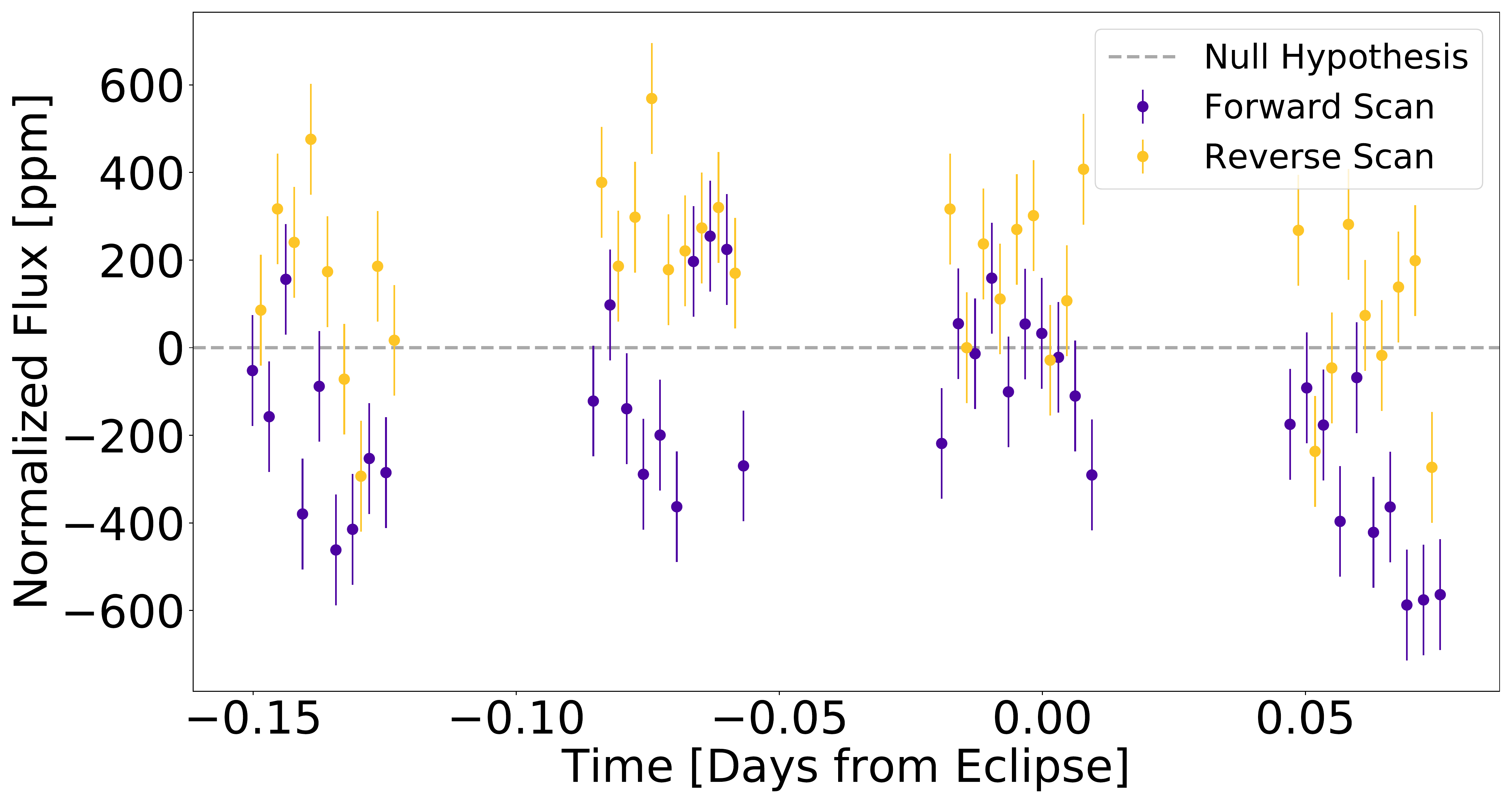}
\caption{\HSTWFCTUVIS{} raw normalized flux light curve with {forward scan values in violet} and  {reverse scan values in orange}. 
There is an apparent difference between the flux measured for forward and reverse scanned traces of 344~ppm.
}
\label{fig:raw_flux}
\end{figure}

\section{Observations}
We photometrically observed the hot Jupiter system \StarName{} during an eclipse event of the planet, which encompassed four \HST{} orbits (see \autoref{fig:raw_flux}). 
This included an initial orbit to mitigate \HST{}'s post-slew positional settling. 
\HST{} observed \PlanetName{} on 2019-07-03 from 00:14:59 UT until 05:40:02 UT,
with our data being processed by the \HST{} Data Processing Software System version \HST{}DP 2019.3.
Below we describe the choices, constraints, and capabilities of \HSTWFCTUVIS{} for observing exoplanets with scanning-mode, visible light photometry.

Our program was one of the first transiting exoplanet programs to use \HSTWFCTUVIS{} in photometric scanning mode. 
{
\citet{Wakeford2020AJ} and \citet{Lewis2020ApJ} recently published spectral-temporal observations of the transiting exoplanet HAT-P-41b using the \HSTWFCTUVIS{}.
They achieved a 29-33ppm precision over the G280 grism spectrum from 200 to 800 nm.
Furthermore, Kenworthy et al (submitted to A\&A; PI: Wang; HST14621 \& HST15119) observed Beta~Pic with \HSTWFCTUVIS{} scanning mode observations to detect the predicted transit of the hill sphere for Beta~Pic b,
and achieved a 57ppm precision photometry from a narrow UV filter.}
Here, we carefully describe our observational program and data analysis procedures to guide further programs and provide greater understanding of our considerations (see \autoref{appendix:arctor} for more details).

\begin{figure}
\centering
\includegraphics[width=0.475\textwidth]{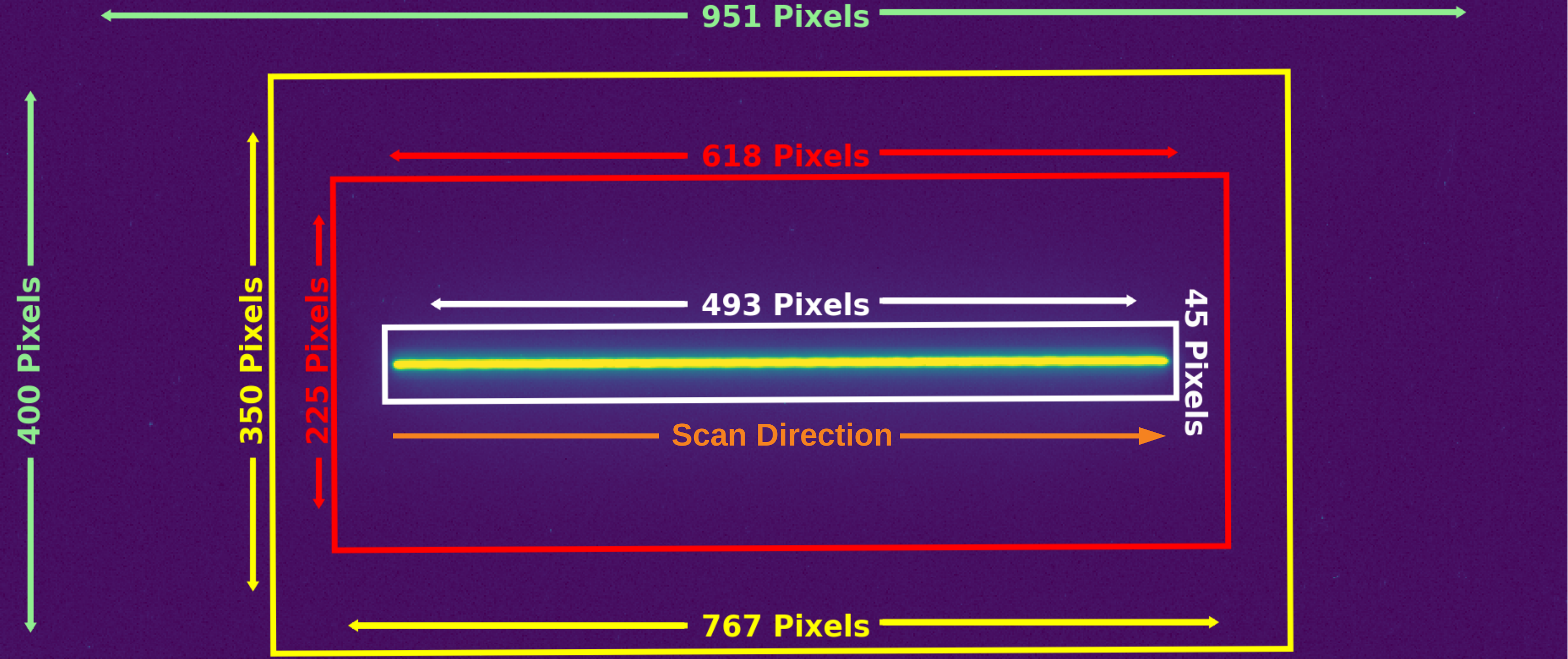}
\caption{\HSTWFCTUVIS{} scanning mode observation example with apertures for photometry and background measurements. All pixels within the white rectangle are summed to measure the flux. The background region consists of the annulus region within the red and magenta rectangles.
The values given represent the 'best' photometry+model pair from our \texttt{AICc} \&  \texttt{BIC} analysis, with a 766 pixel Aperture Width x 45 Aperture Height for our photometry.
The figure shows larger photometry aperture (white) than the 'best' solution pair.
}
\label{fig:image_and_aperture}
\end{figure}

Because \StarName{}, the host star, is moderately bright (V$_{mag} = 12.4$) in our chosen filter (F350LP filter; 346nm - 822nm), we observed it using \HSTWFCTUVIS{} spatial scan mode with a scan rate of 0.2278 arcsec s$^{-1}${; the maximum scanning rate is 1 arcsec s$^{-1}$, limited by flight software.}
\HST{} placed the scan trace perpendicular to the read direction and located 179 pixel rows above the readout edge of the detector (see \autoref{fig:image_and_aperture}).
Each exposure lasted for exactly 82 seconds and spanned 18.68 arcsecs, or 467 pixels.
The expected length of the trace required the use of a custom 400x951 WFC3 sub-array aperture (see \autoref{fig:image_and_aperture}) to maximize the SNR, and ability to estimate the background, while minimizing readout time.

This configuration provided 18-20 frames per orbit (10 forward scan and 10 reverse scan) -- culminating in 75 frames for the entire observation, including the initial orbit which established the precision of our pointing over the eclipse duration.
This configuration resulted in a maximum pixel value of 48000 e-, with an estimated (median) background noise of 15 e- per image.

We used the \texttt{F350LP} filter to maximize throughput and spectral coverage, without saturating the detector. 
{Comparing the photon noise predictions from the official \HSTWFCTUVIS{} ETC} -- with multiple filter/scanning rate/integration time combinations -- our configuration provided the minimum predicted uncertainty per image and maximum number of frames per eclipse.
These observations resulted in a photometric uncertainty on the integrated flux per image of 124~ppm (SNR$\sim$8065); and a global scatter over the raw light curve (i.e.~RMS) of 206~ppm.

\begin{figure*}[!t]
\centering
\includegraphics[width=\textwidth]{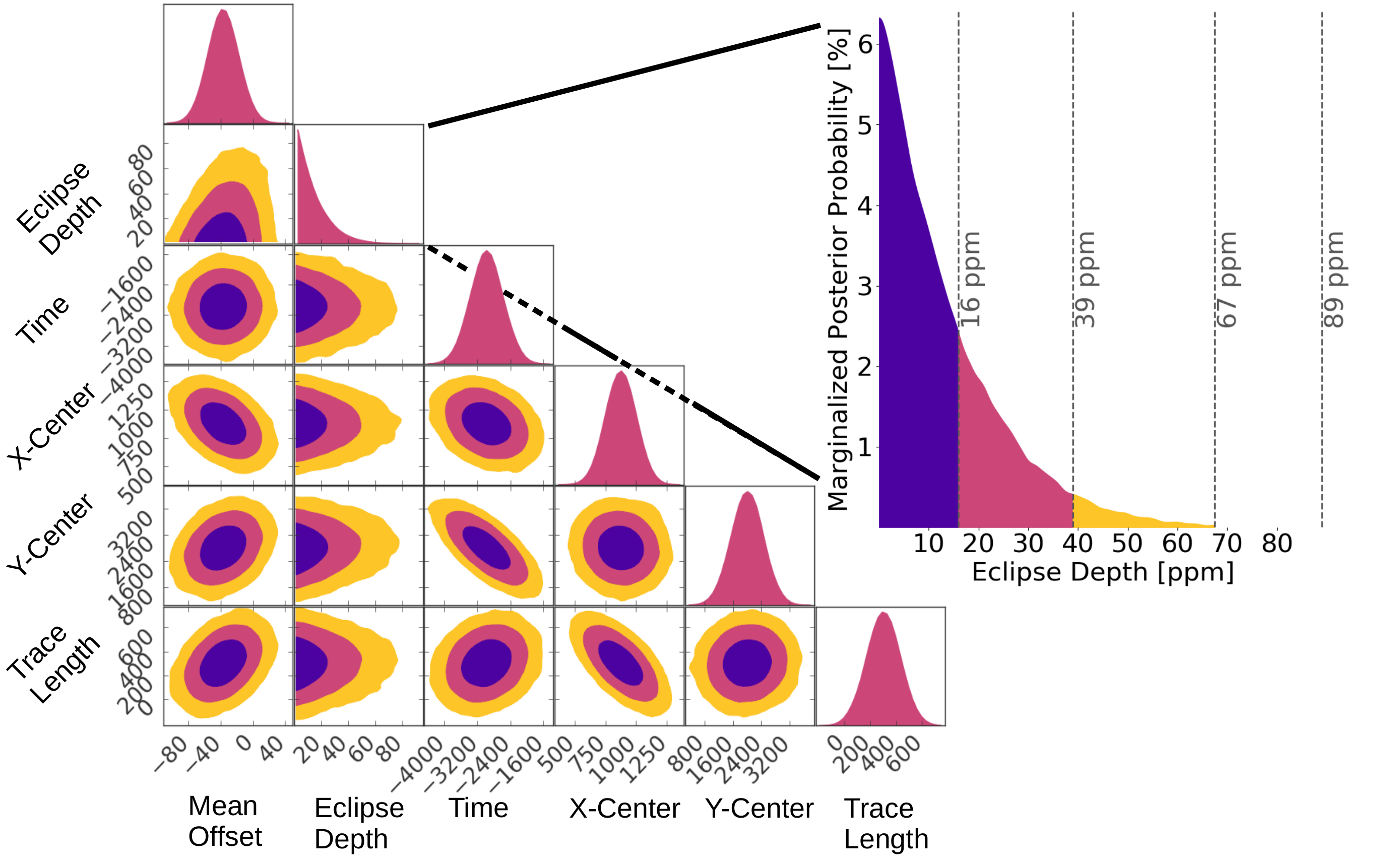}
\caption{Corner plot with Eclipse {Kernel Density Estimates (KDE)}.  The corner plot is a sequential 2D correlation and marginalization description of the MCMC chains. 
The KDE inclusion (upper right) shows a smoothed representation of the histogram over the MCMC posterior for the Eclipse Depth alone. 
All five non-eclipse depth measurements are properly bounded as Gaussian posteriors (see diagonals).
The asymmetric behaviour of the eclipse depth posterior is representative of our non-detection. 
Using quintile analysis, we derived that 3-$\sigma$ (99.7\%) of the chains were sampled within 67~ppm, which sets our upper limit used throughout this paper.
}
\label{fig:corner_plot_best_AIC}
\end{figure*}

\subsection{{Data Reduction}}

{We developed a novel \texttt{Python} package that is optimized for extracting photometric, \HSTWFCTUVIS{}, scanning-mode, time-series observations; this is the first {UVIS} photometric scanning package to be released open-source as a reusable package.}
We named it \texttt{Arctor} because it is optimized to extract photometry from arcs on images; i.e.~traces.
Our software may also be relevant to surveys observing Near Earth Objects that streak across their detectors.
The package is publicly available and BSD3 licensed on our \texttt{GitHub} page\footnote{github.com/exowanderer/arctor}.

\texttt{Arctor} performs standard astronomical, photometric extraction; but it is adapted for scanning mode observations, which spread the light across the detector to maximize the instrument's duty cycle.
The special considerations for scanning mode include: using rectangular apertures for flux; median and column-wise background estimation; as well as several other detailed considerations discussed in \autoref{appendix:arctor}.

After downloading the entire data set from \textit{MAST}
(\texttt{mast.stsci.edu}),
we also extracted the necessary planetary parameters from  \texttt{exo.mast.stsci.edu}
\footnote{{\texttt{exoMAST-API}: github.com/exowanderer/exomast\_api}} \citep{Mullally2019RNAAS}.
For our analysis, we only needed to examine the 75 flat field corrected (FLT) fits files. 

\texttt{Arctor} examined the the positional coordinates (``POSTARGs'') to identify which observations are associated with forward and reverse scanning directions.
We used the time stamps to identify and map temporal correlations with the correlated-noise and astrophysical signals.

Our primary operations were to extract the photometry and scanning trace properties; i.e.~y-positions (cross-scan spatial pixel position), x-positions (read-direction spatial pixel position), trace angles (relative to mid-trace), and trace lengths, as well as the frame indices corresponding to the forward and reverse scanned observations.
Moreover, we also computed the sky background with both static and scanning-focused methods; {we then computed} the cosmic ray mask along the temporal axis (i.e.~statistical outlier rejection in time, per pixel).

\subsection{{Data Analysis}}
After extracting the photometric time-series, we used four, independent model selection methods (i.e.~AICc, BIC) to diagnose whether any of the features listed above were correlated with the flux (see \autoref{fig:corner_plot_best_AIC}).
We used \texttt{Arctor} to fit both MAP (Maximum A Posteriori) solutions and MCMC (Markov Chain Monte Carlo).
{We explored the posterior probability distribution using a Hamiltonian Monte Carlo (HMC)}\footnote{{PyMC3: https://github.com/pymc-devs/pymc3}}$^,$\footnote{{Exoplanet: https://github.com/dfm/exoplanet}} {\citep{Salvatier2016PyMC3, DFM2019ascl}}.
By including our photometric uncertainties ($\sim$124~pmm), we generated our eclipse depth (and other parameters') uncertainties from the BCRs derived by our MCMC posteriors (see \autoref{fig:corner_plot_best_AIC}).

The HMC package (\texttt{exoplanet}) that we used includes a variant of both \text{BATMAN} and \texttt{STARRY} to model transiting exoplanet observations \citep{Kreidberg2015PASP, Luger2019AJ, DFM2019ascl}.
This provided a fully analytic transit light curve, including ingress, egress, and limb-darkening \citep{MandelAgol2002ApJtransits}.
Because the \texttt{exoplanet} package was built for transit modeling, we required that the eclipse depth must be positive.
To establish proper Bayesian posterior bounds, we tested multiple priors: uniform, log-normal, log-uniform; all of which resulted in a non-detection with similar uncertainties over the eclipse depth.
Moreover, we sampled 16 HMC chains for the 25 light curves that sustained the 25 lowest AIC/BIC from the MAP analysis, discussed above.
{Each of our light curves included the eclipse light curve model \citep{MandelAgol2002ApJtransits} and a combination of linear systematic trends with respect to time, x-position, y-position, trace length, and trace angle. We did not explore non-linear systematic trends; but we tested the validity of our final result by including Gaussian Processes \citep{gibson12a, gibson12b, DFM2019ascl} and computing the AIC/BIC with it as well (see below).}
The convergence of each set of 16 chains was confirmed through both inspection of the autocorrelation figure and Gelman-Rubin Test, provided by the \texttt{exoplanet} package.

After deriving our \textit{best} model, we further diagnosed the existence of residual auto-correlated noise sources (i.e.~power-law noise). We diagnosed several techniques: \citet{Carter2009ApJ_wavelets} wavelets, residual binning technique \citep{Pont2006, Cubillos2017AJ}, and Gaussian Processes \citep{DFM2017AJ}. Our MCMC analysis with Gaussian Processes did not recover the existence of significant auto-correlated noise; see appendix-\autoref{gaussian_processes} for further details.

Assuming that no eclipse occurred during our observations, and not accounting for any systematic noise sources -- i.e.~if only Gaussian noise existed -- then we estimated that the minimum eclipse depth that our data could have detected at 3-$\sigma$ would be 102~ppm (i.e.~$\sigma = \frac{206}{\sqrt{75}} \times \sqrt2 \sim 34$~ppm and thus $3 \sigma \sim 102$~ppm).
Our final $3\sigma$ upper limit, after modeling the correlated noise sources, achieved $3\sigma\sim67$~ppm -- a $>$34\% reduction in global noise sources, which reveals the effect of inflated uncertainties from non-modeled correlated noise.

We provide a detailed description of \texttt{Arctor}, our novel pipeline for photometry scanning mode observations in \autoref{appendix:arctor}. 
Furthermore, in \autoref{appendix:map_aicc}, we show the results from our 12800 MAP fits to derive the \textit{corrected}~Akaike~Information~Criterion~(\texttt{AICc}) and Bayesian~Information~Criterion~(\texttt{BIC}) values to select to most viable set of features and hyper-parameters for optimizing the information extraction from our observations. [\texttt{AICc} and \texttt{BIC} are defined in \autoref{appendix:map_aicc}].
We used our 12800 MAP fits to probe the span of all possible combinations of our hyperparameters: linear trends (time, x-position, y-position, trace-length, trace-angle); aperture width and aperture height for our rectangular apertures.

\section{Results}
\begin{figure*}
\centering
\includegraphics[width=\textwidth]{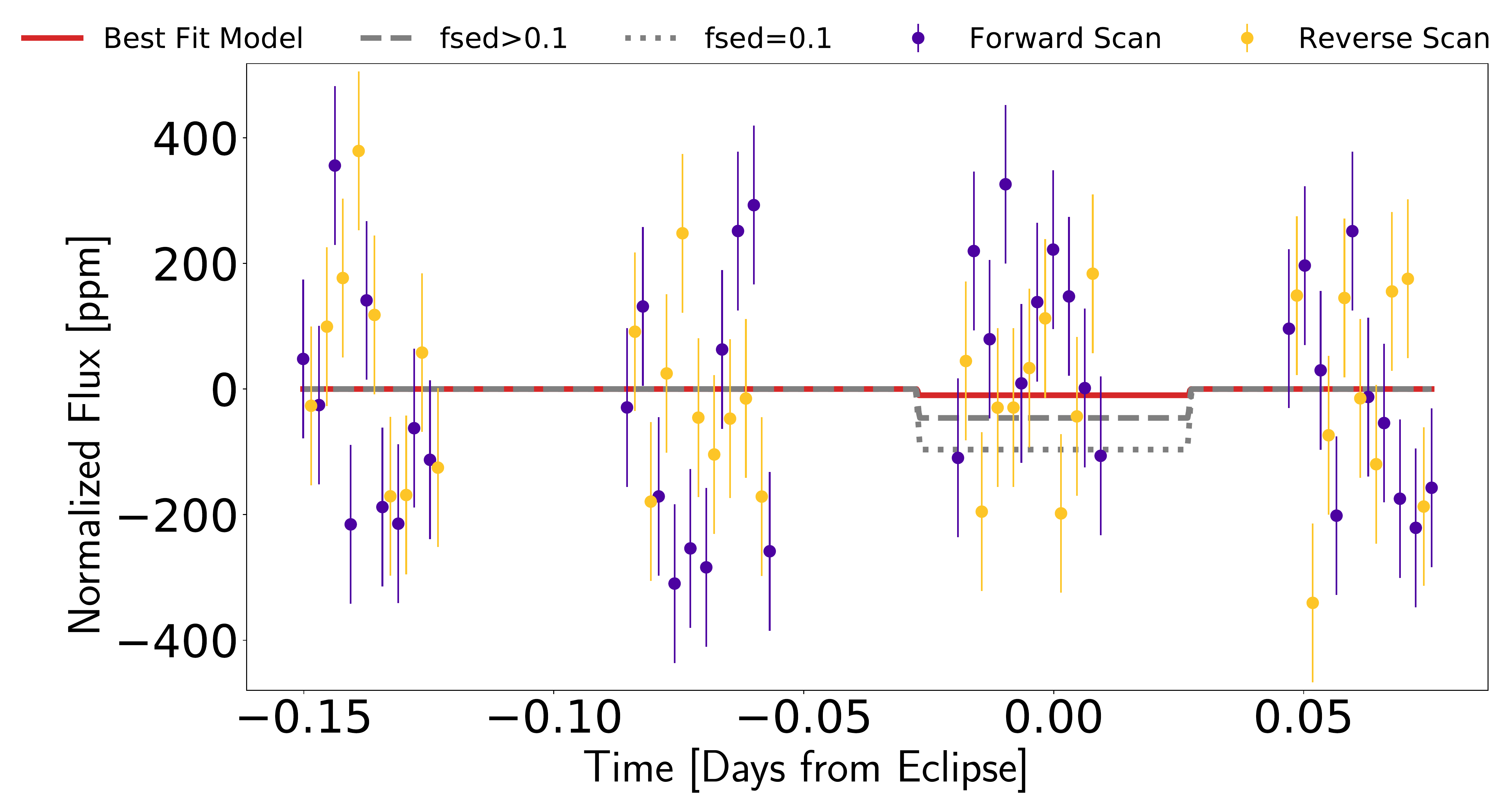}
\caption{\HSTWFCTUVIS{} normalized flux light curve with {forward scan values in violet} and  {reverse scan values in orange}. 
The errorbar points are corrected from fitted systematic behaviour involving linear trends with respect to X-Centers, Y-Centers, and Trace Lengths.
The pale violet and orange background points represent the raw-flux values from our rectangular photometric extraction.
The models shown here represent the null hypothesis of 'no eclipse' (straight black dashed line), our best \texttt{AICc} \& \texttt{BIC} derived model (in red), our predicted eclipse model with \fsed~=~0.1 (grey dashed line), and our predicted eclipse model with \fsed~$>$~0.1 (grey dotted line).
The \fsed~=~0.1 model is not consistent with our observational results, while the \fsed~$>$~0.1 is consistent with our 3-$\sigma$ upper limit.}
\label{fig:best_AIC_flux_model}
\end{figure*}

\begin{figure*}
\centering
\includegraphics[width=\textwidth]{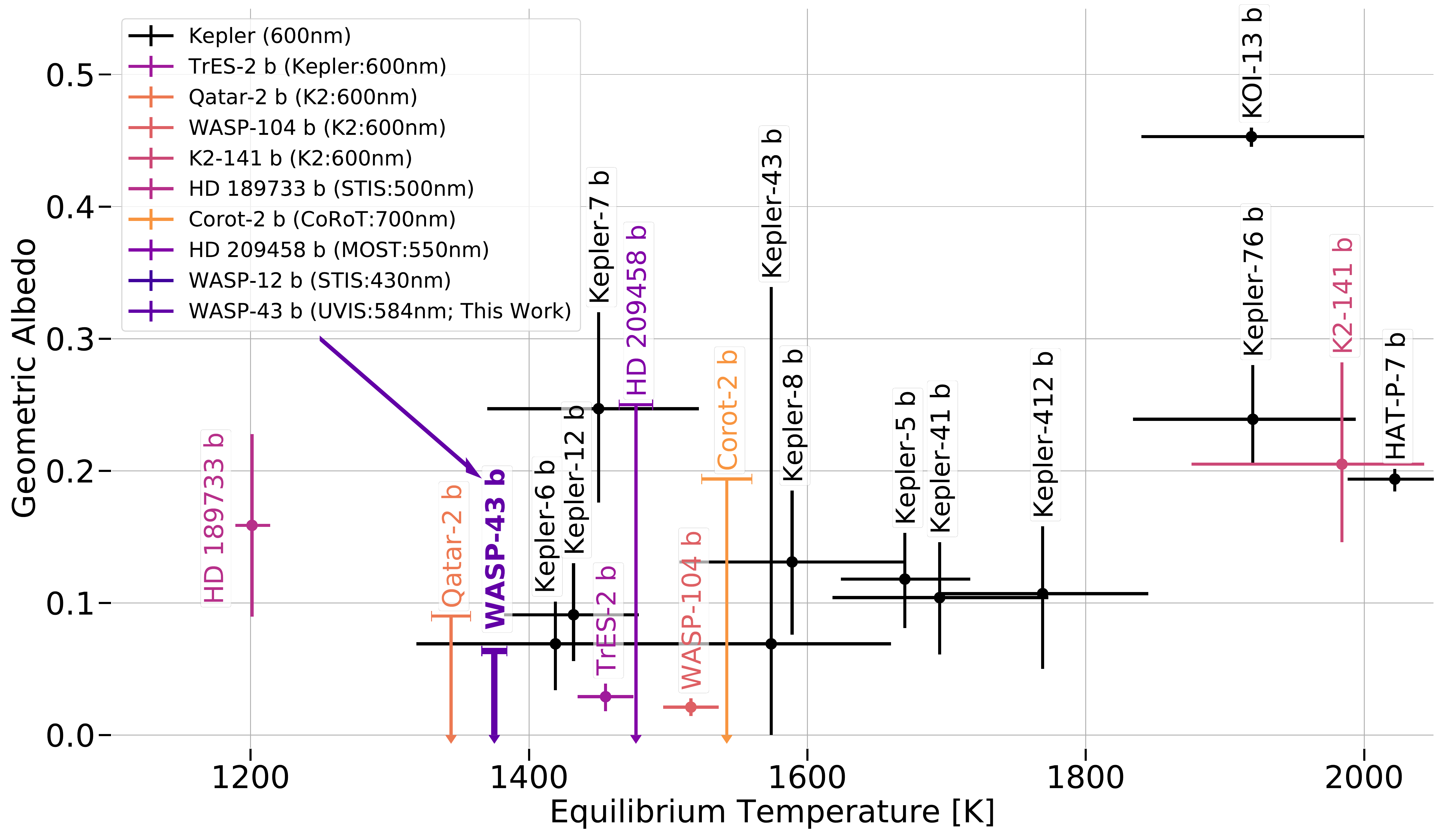}
\caption{Apparent geometric albedos comparisons from similar observations of reflected light eclipses from hot Jupiters. All upper limits are reported here as 3-$\sigma$ upper limits.
The black dots represent results from the \Kepler{} mission \citep{Esteves2015ApJ, Shporer2015AJ, Angerhausen2015PASP, Bell2017ApJ}.
The \Kepler{} points were observed at 600nm, which sustains substantial contamination from atomic absorption contamination and (likely) thermal emission.
The HD 189733b reflected light observations were observed with \HSTSTIS{}, which sustained $\sim2\times$ our uncertainties \citep{Evans2013ApJ, Bell2017ApJ}.
Our \PlanetName{} \HSTWFCTUVIS{} observations sustained the most precise upper limit attained in this wavelength and (equilibrium) temperature range.
}
\label{fig:kepler_comparisons}
\end{figure*}

The overarching result from our observations is that the dayside of \PlanetName{} is very dark. Integrated over the entire optical band (346nm - 822nm), we derived an upper limit to the reflected light eclipse depth of \textit{less than} 67~ppm, corresponding to a geometric~albedo~of~A$_\text{g} \lesssim$~0.06 (3$\sigma$ upper limit); see \autoref{fig:corner_plot_best_AIC},~\ref{fig:best_AIC_flux_model},~and~\ref{fig:kepler_comparisons}.

Our individual frames sustained an average 124~ppm photometric uncertainty, which was $<$1\% different from the predicted photon+read noise using the \HSTWFCTUVIS{} ETC. This resulted in our \textit{best} AICc parameter set generating a light curve with the Standard Deviation of the Normalized Residuals (SDNR)~=~172~ppm; and a model uncertainty on the eclipse depth of 34~ppm.

\autoref{fig:best_AIC_flux_model} shows the best fit eclipse depth (9~$\pm$~34~ppm), {along with other predictive eclipse depths for relevant cloud  models (see below).}
Our lack of detection of an optical eclipse depth implies a lack of reflective clouds on the dayside, within the layers probed by eclipse observations (P~$\lesssim$~1~bar).

Our derived upper limit of 67~ppm corresponds to a geometric albedo, $A_g \lesssim $ 0.06.
\Kepler{} (and subsequently K-2) detected the optical eclipse of more than a dozen hot Jupiters \citep{Esteves2015ApJ}.
\autoref{fig:kepler_comparisons} displays their geometric albedos, as a function of equilibrium temperatures (T$_{\text{eq}}$), 
which includes several detected reflected light eclipses in a similar wavelength range ($\sim$600~nm) to our \HSTWFCTUVIS{} F350LP observations. 
Compared to this set of \Kepler{} reflected light results, our non-detection would imply the atmosphere is less likely to sustain clouds (in the observational regime) than planets with similar T$_{\text{eq}}$; i.e.~within~$\sim$100K of \PlanetName{} \citep{Esteves2015ApJ, Niraula2018arXiv}.

\autoref{fig:kepler_comparisons} places our reflected light geometric albedo upper limit within the context of many known, high-precision measurements with 1000~K~$<$~T$_{\text{eq}}$~$<$~2000~K -- all from space-based facilities.
Our observation is not the highest precision geometric albedo constraint, but it is equivalent to observations with multiple epochs; i.e.~TrES-2 b and WASP-104b -- both observed with \texttt{K2}.
In contrast, it is the most precise geometric albedo upper limit, and at the second lowest T$_{\text{eq}}$.

The distribution of geometric albedos over equilibrium temperatures shown in \autoref{fig:kepler_comparisons}, reveals an apparent linear trend, such that hotter planets have a larger geometric albedo. 
On the other hand, given the limited data, large uncertainties, and  possible thermal contamination for the hottest planets, this trend is not conclusive.

Comparing to the thermal infrared phase curve results  (see \autoref{fig:toy-model}), \PlanetName{} does appear to have an anomalously large thermal day-to-night contrast \citep{Stevenson2017AJ}, implying a lack of significant atmospheric circulation; {combined} with the apparent lack of observable clouds on the dayside of the planet, this could imply that the thermal phase curve observations are the result of different pressure levels being observed at the day~side, as compared to the limbs and night~side of the planet (i.e. maintaining $\tau\sim1$) \citep{Stevenson2014Science, Kataria2015ApJ, Stevenson2017AJ}.

\section{Modeling}
\begin{figure*}
\centering
\includegraphics[width=\textwidth]{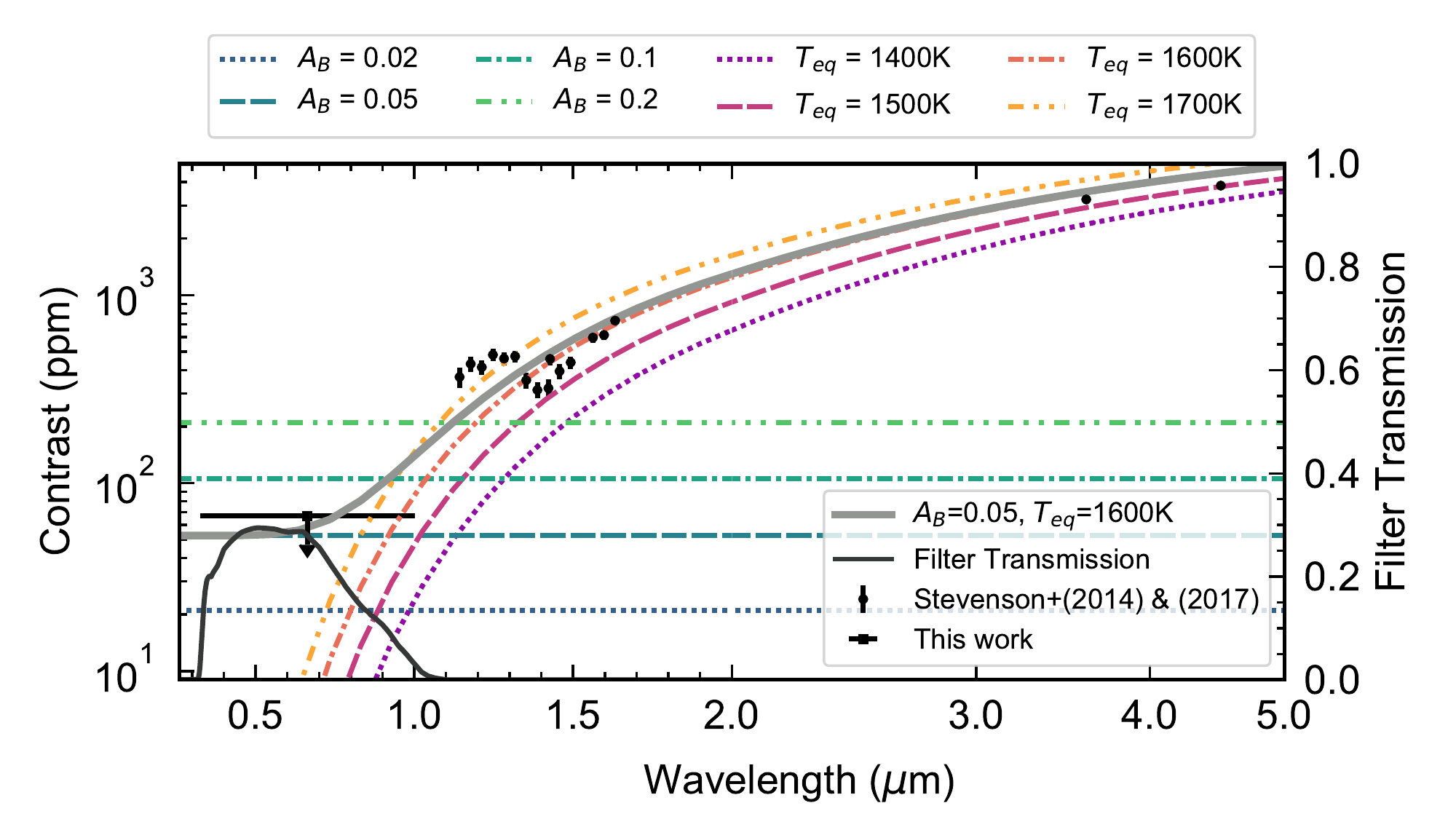}
\caption{Toy model example of what Bond albedos and equilibrium temperatures are allowed by the \HST{} observations presented here and the \HST{} WFC3 data of \citet{Stevenson2014Science} and the \Spitzer{} data of \citet{Stevenson2017AJ}. In horizontal line, we show $A_B$ = 0.02, 0.05, 0.1, 0.2 along with several blackbodies in shades of purple to orange of $T_{\rm eq}$ = 1400, 1500, 1600, and 1700K. The example combination of $A_B$ = 0.05, $T_{\rm eq}$ = 1600 K allowed by the data is shown in gray, implying that the planet is very dark on the dayside.}
\label{fig:toy-model}
\end{figure*}

The eclipse depth indicates that the reflected light and/or thermal emission from the planet is very low over this NUV/Optical band~pass. The predicted dayside temperature of the planet ranges between 1400~K and 1600~K at observable wavelengths \citep{Kataria2015ApJ, Stevenson2017AJ}. Thus, in order to produce a suitably low eclipse depth, we can predict a maximum allowable albedo in the optical photometric band. \autoref{fig:toy-model} shows a toy model schematic of potential possibilities. The model combines the reflected light signal expected from a planet with a monochromatic Bond albedo (the fraction of the total light reflected compared to incident stellar light) of 0.02, 0.05, 0.1, or 0.2 with the thermal emission expected from a blackbody of 1400~K, 1500~K, 1600~K, or 1700~K. Given the eclipse depth upper limit of 67~ppm, the planet must be very dark and we can expect the planet to have a Bond albedo of 0.06 or less on its dayside hemisphere.

Previous studies of \PlanetName{} predict Bond albedos of $0.36^{+0.11}_{-0.12}$ \citep{Schwartz2015}, and 0.3$\pm$0.1 \citep{Keating2017ApJ} using day and night temperature differences, or geometric albedo conversions, which are ultimately an assumption about the bolometric flux from around the planet; such as the technique used by \citet{Stevenson2017AJ} to measure $0.19^{+0.08}_{-0.09}$.
Each technique is limited by our ability to assume the correct spectral energy distribution (SED) of the planet, as well as the stellar SED.
General Circulation Models (GCMs) can provide a framework to explore the planet holistically.

\begin{figure}
\centering
\includegraphics[width=\linewidth]{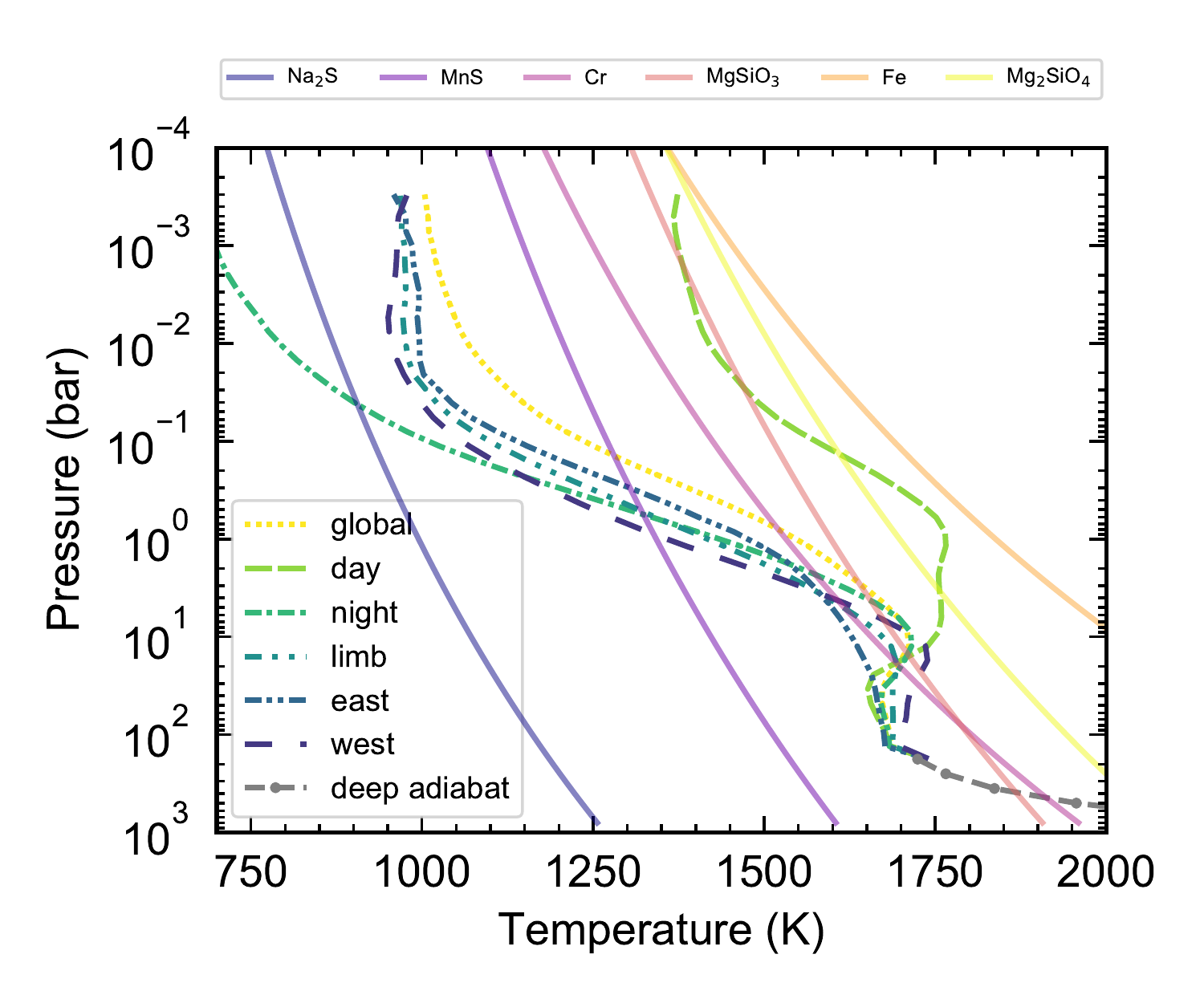}
\caption{The temperature-pressure profiles from \citet{Kataria2014ApJ} along with the condensation curves of several cloud species \citep{Morley2012ApJ}. If the convective region deep in the atmosphere is hotter, Cr and the Mg species will likely only condense in the deep interior. Thus we expect for Na$_2$S and MnS to be the only species available at the pressure probed by our observations}
\label{fig:tiff-tp}
\end{figure}

We explored the possible atmospheric scenarios using pressure-temperature files and the associated vertical mixing profiles ($K_\mathrm{zz}$) generated in the GCMs from \citet{Kataria2014ApJ}, which assume a cloudless atmosphere in chemical equilibrium. {To compare our results with published \PlanetName{} atmospheric models (i.e.~\citealp{Kataria2015ApJ}), we used the same planetary and stellar parameters of \citet{Hellier2011AA}}; which can be found in \autoref{tbl:params}. The temperature-pressure profiles are shown in \autoref{fig:tiff-tp}, over-plotted on condensation curves for several cloud species. Starting from the deep interior, as the planet's upper layers decrease in temperature, a particular cloud species may condense with a cloud base at the intersection of the condensation curve and the temperature-pressure profile. 

\begin{deluxetable}{ll}
\tablecaption{Planetary and stellar parameters assumed in atmospheric modeling. \label{tbl:params}}
\tablehead{\colhead{Parameter} & \colhead{Value}}
\startdata
$P$ & 0.81347~days \\
$a$ & 0.01526~AU \\
$i$ & 82.33\degr{} \\
$M_\mathrm{p}$ & 2.05~M$_{\rm J}$ \\
$R_\mathrm{p}$ & 1.036~R$_{\rm J}$ \\
$g$ & 47.342~m\,s$^{-2}$ \\
$M_\mathrm{*}$ & 0.717~M$_\odot$ \\
$R_\mathrm{*}$ & 0.667~R$_\odot$ \\
$T_\mathrm{eff}$ & 4300.0~K \\
\enddata
\end{deluxetable}

We used the cloud code developed by \cite{Ackerman2001} to compute cloud profiles derived from temperature-pressure profiles from \cite{Kataria2014ApJ}.
We allow Na$_2$S, MnS, Cr, and MgSiO$_3$ to condense. We also test five sedimentation efficiencies, \fsed{}, which controls the particle size and vertical extent of the cloud. A smaller \fsed{} yields clouds which are vertically extensive while larger \fsed{} values yield vertically thin clouds. We then used \verb|PICASO| \citep{Batalha2019ApJ} to generate the geometric albedo spectrum of each case (36 cases); i.e.~when the full phase dayside appears at secondary eclipse. Note that the geometric albedo spectrum is defined to be at full phase. The night-side average spectrum would never be observed in such a geometry, but is illustrative to show the variations in cloud structure inferred around the planet. We used the opacity database supplied with the initial public release and a PHOENIX model representation of \StarName{} \citep{Batalha2019Zenodo, pysynphot2013ascl}.

\begin{figure*}[!ht]
\centering
\includegraphics[width=\linewidth]{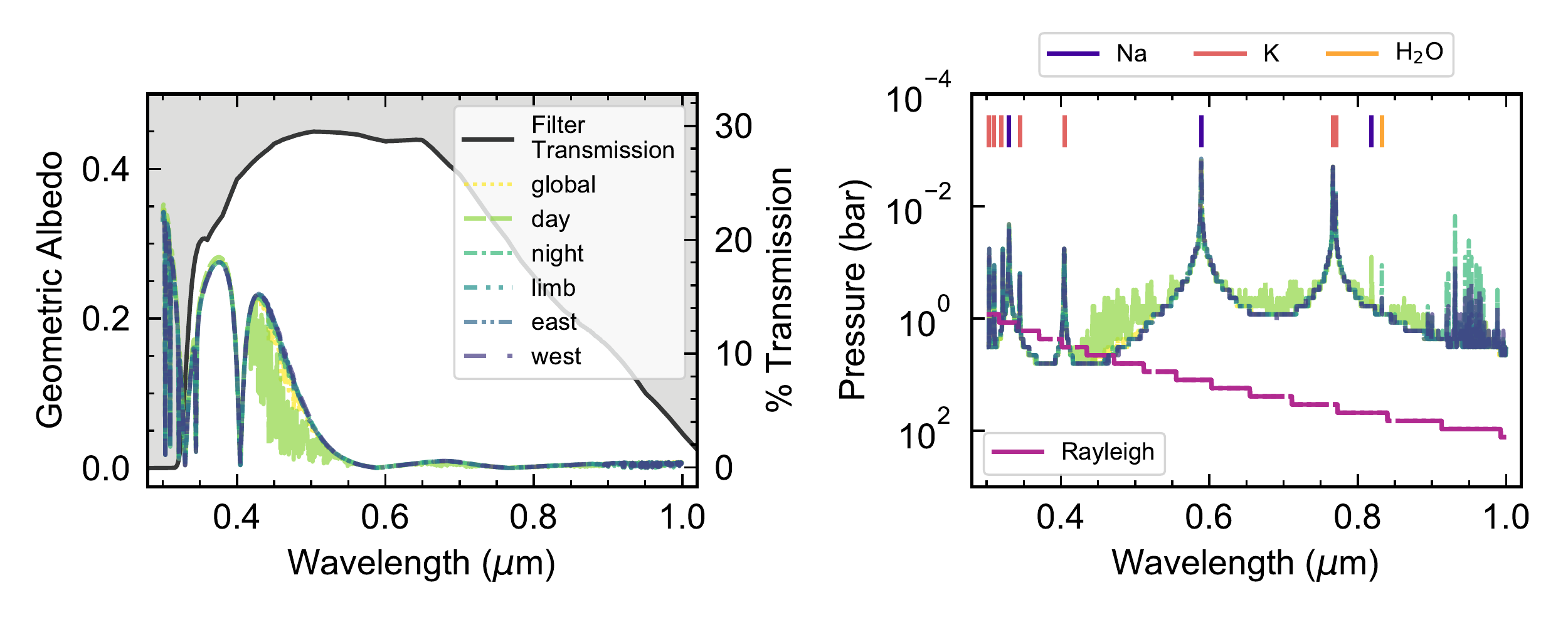}
\caption{The geometric albedos for the cloudless cases. Left: the geometric albedo for each of the six profiles. Right: the photon attenuation diagram. The lines correspond to the pressure level at which the two-way optical depth in the atmosphere reaches $\tau$=0.5, caused by the gas absorption (colors and linestyles corresponding to the left panel) and cloud or Rayleigh scattering. Without clouds the gas chemistry variations between the GCM averages produce the observed albedo spectrum variations. Lines are slightly transparent to demonstrate how similar the gas chemistry is among all other profiles. We mark the positions of prominent Na, K, and H$_2$O (on the night-side model only) absorption lines. The less prominent lines near 0.5\,$\mu$m and 0.7\,$\mu$m and 0.9\,$\mu$m are due to TiO and VO in the dayside model. The night-side instead shows H$_2$O and CH$_4$ lines beyond 0.9\,$\mu$m.}
\label{fig:cloudless}
\end{figure*}

The geometric albedos for the cloudless cases are shown in the left panel of \autoref{fig:cloudless} and can be understood by examining the photon attenuation diagram in the right panel. The lines correspond to the pressure level at which the two-way optical depth in the atmosphere reaches $\tau$=0.5, caused by the gas absorption and cloud or Rayleigh scattering. Without clouds, albedo differences are driven by the gas chemistry with the largest features being caused by Na and K, as well as water at longer wavelengths. The hotter dayside (dotted lines) shows additional gas opacity lines (such as near 0.5~$\mu$m), which are caused by TiO and VO; this is reflected in a much darker albedo spectrum in those wavelength regions, but is a weak effect.

\begin{figure*}
\centering
\includegraphics[width=0.85\textwidth,keepaspectratio]{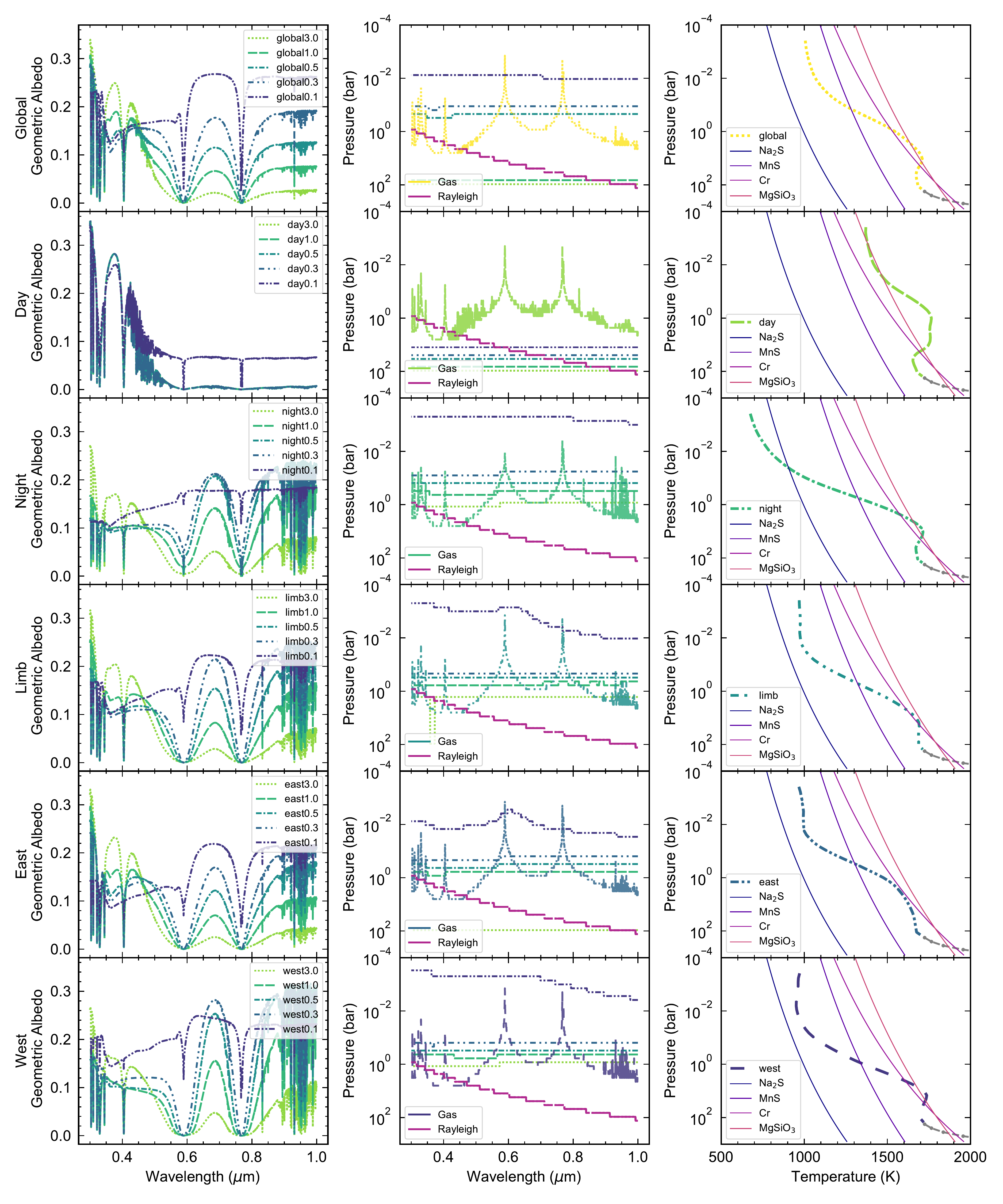}
\caption{The geometric albedos for the cloudy cases. Each row is a different profile from \citet{Kataria2015ApJ}. Left: the geometric albedo spectrum for each of the five sedimentation efficiencies. Middle: the photon attenuation diagram showing the pressure level at which the two-way optical depth in the atmosphere reaches $\tau$=0.5, caused by the gas absorption and cloud or Rayleigh scattering. Linestyles and colors for the cloud opacity correspond to their geometric albedo spectrum counterparts. Note the pressure level where clouds contribute changes with \fsed. Right: the temperature-pressure profile overlaid on the cloud condensation curves. All profiles readily produce clouds of which only the most lofted begin to dominate as an opacity source except for the dayside profile.}
\label{fig:cloudy}
\end{figure*}

Once clouds are added, the location of the cloud varies across the planet and how high the cloud particles are lofted begins to affect the albedo spectrum.
In \autoref{fig:cloudy}, we show the results of the cloudy cases where Na$_2$S, MnS, Cr, and MgSiO$_3$ are allowed to condense at depth.
From top to bottom are the different sedimentation efficiency cases for: global-average, dayside, nightside, limb-average, and east or west limb (the temperature-pressure profiles are shown in the right panel). In nearly all profile scenarios, the \fsed=3.0 cloud scenario is very similar to the cloudless case presented in \autoref{fig:cloudless}. The shape is then modulated by continuing to loft the clouds higher and higher into the atmosphere such that the cloud begins to dominate over Rayleigh scattering and gas absorption, leading to a darker albedo spectrum at short wavelengths, but also increases the reflectivity at long wavelengths.

In agreement with the previous conclusions of \citet{Kataria2015ApJ}, we predict that the nightside should readily produce clouds (the upper most is Na$_2$S). These clouds should be high enough in the atmosphere that, at various sedimentation efficiencies, they make a significant impact on the albedo spectrum; the effect is to darken the spectrum at shorter wavelengths and brighten it at longer wavelengths.
For smaller particles, Na$_2$S gets darker beyond 1~$\mu$m, but for larger particles it predominantly remains at the same brightness.
The pressure at which the cloud(s) exist will determine the observed brightness temperature and albedo. Therefore, the observed brightness temperature can constrain their sedimentation efficiency and, related, particle sizes.
The most significant effect is that a MgSiO$_3$ cloud layer would serve to dramatically brighten the albedo spectrum.

On the dayside, spectroscopy may be capable of determining the presence of TiO and VO in the gas phase by determining the albedo near 0.5~$\mu$m. The model is unable to condense a cloud at observable pressures on the dayside, unless MgSiO$_3$ is included; and even then, only for the smallest \fsed.
In the smallest sedimentation efficiency case some small particles are mixed up, through the atmospheric temperature inversion in the dayside profiles, thereby delivering condensible gas to higher regions of the atmosphere where it can once again condense. Note that \PlanetName{} has an exceptionally high gravity (g~$\sim$~51~m$\cdot$~s$^{-2}$) compared with other hot Jupiters like Kepler-7b (g~$\sim$~4~m$\cdot$~s$^{-2}$).

As a result, in the absence of particles representing a small \fsed, the albedo spectrum would be dark, which is consistent with the upper limit derived from the data; this would imply that it is entirely dominated by gas absorption. More aggressive vertical mixing schemes are able to deliver more MgSiO$_3$ particles to observable pressure, but the particles grow in size, which are less reflective. Whether or not such a second cloud layer would form in a realistic atmosphere, as opposed to the idealized case studied here, would take more sophisticated modeling.

The models and data continue to suggest that \PlanetName{} has inhomogeneous cloud coverage, where the dayside has essentially no observable clouds, while the limbs and night sides are able to condense all four cloud species.
Typically the upper most cloud is MnS, except on the nightside where the models predict Na$_2$S to be the upper most cloud.
In contrast, if the bright MgSiO$_3$ particles were comparatively smaller they become more difficult to sequester with depth. The subsequent cloud layer would extend to the top of the atmosphere and become the dominant source of cloud opacity at all planetary longitudes.

\begin{figure*}
\includegraphics[width=\linewidth]{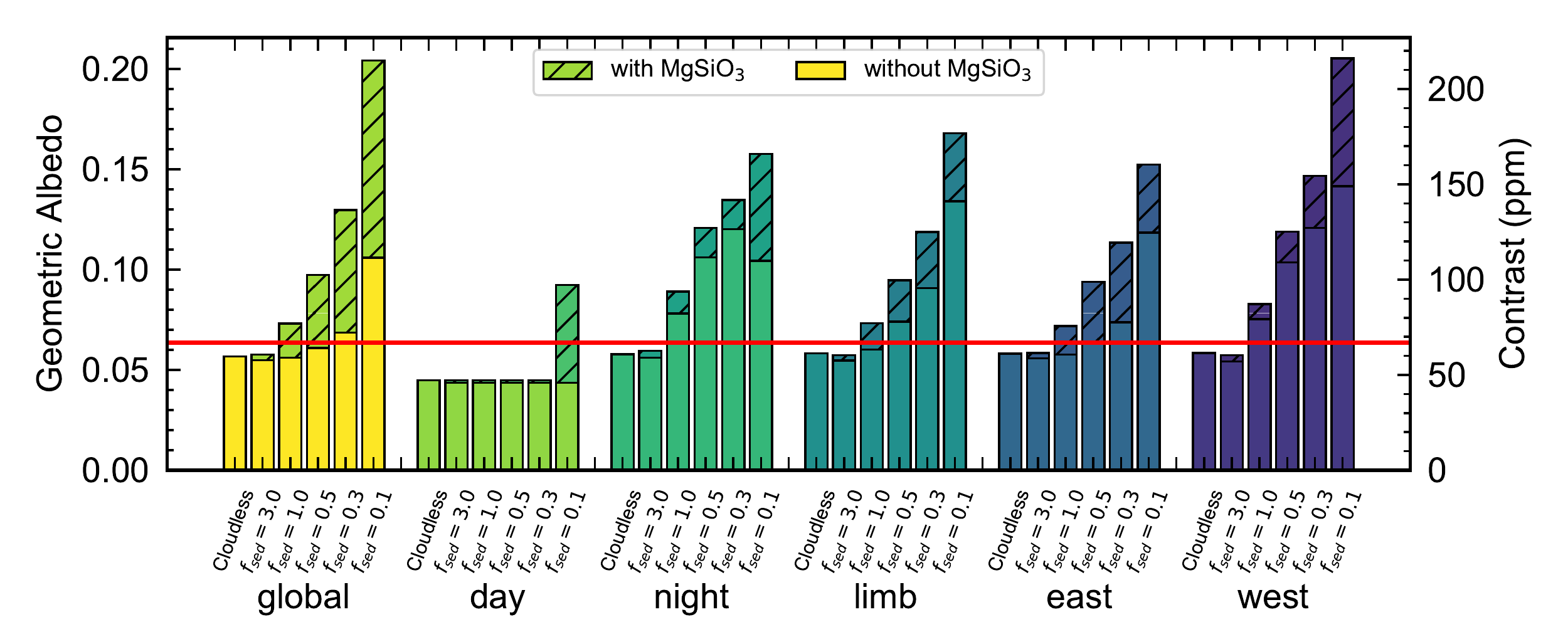}
\caption{The filter-integrated geometric albedos of all model cases excluding MgSiO$_3$ condensates (solid bars) and including (hatched bars).
Our measured constraint on the observational 3-$\sigma$ limit (in contrast) is represented by the horizontal red line.
Our upper limits imply that the \fsed=0.1 dayside model is not consistent with our data; we infer from this that either small lofty particles are not being propagated into the upper atmosphere (i.e.~the observable regime; P~$\lesssim$~1~bar), or that cloud forming particles are sequestered in the deep reaches of the planetary atmosphere.
Connecting this to the condensation curves in \autoref{fig:tiff-tp}, we can infer that MgSiO$_3$ -- which thermochemical equilibrium would predict to form clouds at P~$\lesssim$~1~bar -- may have formed clouds within the deep adiabat, thus significantly reducing the available mass to form clouds near our observational regime.
}
\label{fig:modelcontrasts}
\end{figure*}

From each model, we filter-integrate the albedo spectra and compute the expected flux contrast. \verb|PICASO| is also able to compute the thermal emission shortward of 1~$\mu$m and the expected thermal contamination is negligible. These contrasts are shown in \autoref{fig:modelcontrasts}. The dayside is consistent with a null detection of a secondary eclipse. The data rule out the presence of highly lofted MgSiO$_3$ clouds. If it did condense in the dayside atmosphere, only the most lofted scenario (forming a second cloud deck near 10~mbar) would have been detectable. Our models suggest that the pressures being probed by nightside and dayside observations are drastically different because the clouds are more significantly lofted on the nightside, relative to the dayside.

\section{Discussion}

Our modeling results (described in \S4) revealed that to sustain this non-detected reflected light eclipse, the dayside atmosphere must significantly lack particles lofted into the observable part of the atmosphere. The models suggest that only the most lofted cloud case scenario, \fsed=0.1, could escape a cold-trap at the deepest layers of the atmosphere to condense into a second upper cloud layer, which could lead to a detectable optical eclipse depth. All other of our modeled sedimentation efficiencies led to particles being trapped at deeper pressures; they were thus unable to be mixed upwards and re-condensed. The fact that we did not detect an optical eclipse confirms that we can rule out a high altitude, bright, uniform cloud layer.

At planetary longitudes away from the sub-stellar point, the upper atmosphere of the planet is predicted to be cool enough to condense other species, notably, Cr, which is also potentially sequestered deep in the atmosphere; MnS; and, on the nightside, perhaps even Na$_2$S. If the cloud behaviour can be parameterized with \fsed{}\textless 0.5, then these species may lead to an observable phase curve signature -- especially at optical wavelengths. \fsed{} values less than 1 have been increasingly necessary to describe the observations of hot Jupiters such as Kepler-7b \citep{Demory2013, Webber2015ApJ}. If MgSiO$_3$ is not sequestered at depth, or simply not lofted high enough into the upper atmosphere, then the planet would be brighter at other observed phase angles where the temperatures are cooler, allowing larger particles to more easily lofted to form clouds at observable pressure ranges (see again \autoref{fig:modelcontrasts}). If silicate clouds are trapped in the deep convective layers of the atmosphere, then our observations would be unlikely to observe them at lower pressures \citep{Powell2018ApJ}. On the other hand, Cr has a deeper cloud base in the dayside profile; this could thus require more vigorous vertical mixing, with even smaller particles, to reach the upper altitudes where our observations might be able to detect their reflected light.

Determining the parameters for clouds in the atmosphere of \PlanetName{} may constrain whether this planet follows in the footsteps of its brown dwarf cousins or, instead, low gravity hot Jupiters. The presence of a partially cloudy (i.e.~patchy) daysides for planets spanning a wide range of equilibrium temperature led \cite{Parmentier2016ApJ} to conclude that the cloud composition must vary with the equilibrium temperature of the planet. This would imply a transition in cloud composition similar to the L/T transition in brown dwarf atmospheres. With an equilibrium temperature of 1450K, \PlanetName{} is close to this proposed boundary for such a planetary L/T transition, but with a higher gravity than many of its peers. The precise equilibrium temperature at which the trapping mechanism can become effective depends on the species in question and the deep thermal structure of the planet, which is unconstrained by existing observations \citep{Thorngren2019ApJ}.

Because of its observational and atmospheric viability for spectroscopic detections, \PlanetName{} has become a benchmark planet for current and future hot Jupiter observations. Upcoming observations by \JWST{} for both ERS (PI:~Batalha) and GTO (PI:~Birkmann) include $\sim$48 hours of \JWST{} observations to map the thermal structure and chemical composition of this exoplanet with exquisite detail \citep{Bean2018PASP, Venot2020ApJ}. We expect that no other exoplanet has or will be observed with this much precision and wavelength coverage for many years to come. And yet, all of these observations probed, or will probe, the atmosphere at $> 1 \mu m$; in contrast, optical light ($< 1 \mu m$) is the primary component of the atmospheric energy budget that is a direct probe of cloud distributions, particle size, and composition. 

Our lack of understanding about cloud properties -- effective height composition, patchiness at the limb, particle size -- is currently a significant limitation to measure precise molecular abundances in hot Jupiter atmospheres \cite{Ormel2019AA}. With \PlanetName{}, \citet{Kataria2015ApJ} hypothesized that inhomogeneous clouds could explain the lack of nightside flux from \PlanetName{}, as observed in both \HSTWFCTIR{} and \Spitzer{} phase curves \citep{Stevenson2014Science, Stevenson2017AJ}. The presence of such inhomogeneous clouds would increase the longitudinal brightness contrast of its atmosphere -- cloudy at the limb, clear near the sub-stellar point. The extent, composition, and distribution of aerosols in exoplanet atmospheres is one of the most significant standing questions in exoplanet characterization research. Not accounting for inhomogenous clouds could lead to biases in atmospheric abundance retrievals and even more spurious molecular detections when the exoplanetary spectrum is interpreted with independent one-dimensional models at multiple phases, which is usually the case \citep{Feng2016}. Our \PlanetName{} optical eclipse measurement is conclusive that clouds are not uniform, nor bright; but to distinguish between several other implications of our result, we would need a full phase curve of \PlanetName{} in at NUV or Optical wavelengths.
\section{Conclusions}
We observed \PlanetName{} during secondary eclipse with four \HST{} orbits using the \HSTWFCTUVIS{} channel in scanning mode. 
Because this is one of the first transiting exoplanet programs to use \HSTWFCTUVIS{} in photometric scanning mode, we described our observational program and data analysis procedure in detail (see \autoref{appendix:arctor}).
We created a new, multi-functional code for Arc-photometry, called \texttt{Arctor}\footnote{https://gihub.com/exowanderer/Arctor}, which can be used for high-precision photometry of all \HSTWFCTUVIS{} scanning mode observations; and may also be useful for Near Earth Object streak observations.
Using the F350LP filter, we did not detect an eclipse of \PlanetName{} during our observations.
As a result, we robustly constrained an upper limit on the eclipse depth of 67~ppm with 3$\sigma$ confidence.
We determined this upper limit by integrating under the {marginalized uncertainty distributions}, as seen in \autoref{fig:corner_plot_best_AIC}, up to 99.7\% posterior probability.

This upper limit on the eclipse depth is consistent with a very low dayside geometric albedo ($<0.06$) for \PlanetName{}. Using a combination of three-dimensional general circulation and cloud condensation models, we estimated the range of planet-to-star contrasts expected under various thermal and cloud sedimentation efficiency regimes. We find that our 
measured upper limit on the visible wavelength eclipse depth of \PlanetName{} is inconsistent with significant cloud coverage on the dayside of the \PlanetName{}. This finding indicates that if clouds are forming in other regions of the planet's atmosphere, such as on the nightside or at higher pressures, they are not being efficiently transported into the portions of \PlanetName{}'s dayside atmosphere probed by the observations presented here. 

\PlanetName{} is one of the best studied exoplanets to date, with multi-wavelength transmission, emission, and now reflected light observations spanning the visible to infrared. Its relatively high average gravity and short orbital period make it, thus far, a unique laboratory for testing how these planetary properties shape atmospheric physics. Further observations are needed to better understand the three-dimensional thermo-chemical structure of this intriguing planet. There are currently planned phase-curve observations of \PlanetName{} by \JWST{} with GTO~1224 (PI:~Birkmann) and ERS~1366 (PI:~Batalha) \citep{Bean2018PASP, Venot2020ApJ}.
These future observations will probe the infrared emission of \PlanetName{} as a function of orbital phase from 2.5 to 12 {\micron}. 
They are expected to provide an exquisite look into \PlanetName{}'s atmosphere, but may not provide a complete picture of processes like cloud formation.
Our results and models show that further observations of \PlanetName{} at optical wavelengths, with narrower bandpasses or spectroscopy, at multiple orbital phases are needed to better understand the processes shaping \PlanetName{}'s low dayside geometric albedo.  

\section*{Acknowledgements}
This work is based on observations made with the NASA–European Space Agency {\it Hubble Space Telescope} that were obtained at the Space Telescope Science Institute (STScI), which is operated by the Association of Universities for Research in Astronomy, Inc., under NASA contract GO-15473. 
The authors thank Mark Marley for discussions that greatly improved the modeling efforts of this manuscript.
Work performed by L.~C.~M. was supported by the Harvard Future Faculty Leaders Postdoctoral fellowship.
This work could not have been completed without the fundamental opacity work by Richard Freedman and Roxana Lupu.
The authors are grateful to Drs.~Dan Foreman-Mackey and Eric Agol for detailed conversations about auto-correlated noise, Gaussian Processes, and fundamental statistics.
The research leading to these results has received funding from the European Research Council (ERC) under the European Union’s Horizon 2020 research and innovation programme (grant agreement no. 679633; Exo-Atmos).
This research has made use of the NASA Exoplanet Archive, which is operated by the California Institute of Technology, under contract with the National Aeronautics and Space Administration under the Exoplanet Exploration Program.
This project made use of the exo.mast\footnote{exo.mast.stsci.edu} web service (and API) hosted by the Mikulski Archive for Space Telescopes (MAST) at STScI  \citep{Mullally2019RNAAS}.
STScI is operated by the Association of Universities for Research in Astronomy, Inc.
{This research was fostered by members of the Space Telescopes Advanced Research Group on the Atmospheres of Transiting Exoplanets (STARGATE) Collaboration.}

\facilities{{\it Hubble Space Telescope}, MAST, NASA Exoplanet Archive}
\software{python, numpy, astropy, scipy, matplotlib, pandas, picaso, pymc3, exoplanet, exomast\_api, autograd, theano}

\appendix


\section{Arctor \HSTWFCTUVIS{} Scanning Mode Photometry Pipeline Configuration}
\label{appendix:arctor}
Here we describe our novel pipeline, \texttt{Arctor}, which is optimized to extract exoplanet transit \& eclipse light curves from \HSTWFCTUVIS{} photometric light curves, in scanning mode. It may also be useful for Near Earth Object flux estimates for \textit{streak} observations.
\texttt{Arctor} performs similar tasks to standard aperture photometry, except that it is optimized for scanned images, which leave a \textit{trace} on the detector, instead of a PSF.
{\texttt{Arctor} is novel because it the first photometric scanning mode pipeline to be released for open-source development and use; it implements several key algorithms that are not necessary for non-scanning mode observations; and implements new Bayesian inference techniques that have not been included in other transiting exoplanet pipelines.}

After downloading the entire data set from  \texttt{mast.stsci.edu}, we also extracted the necessary planetary parameters from  \texttt{exo.mast.stsci.edu}\footnote{github.com/exowanderer/exomast\_api}  \citep{Mullally2019RNAAS}.
We examined the 75 provided FLT files (FLT represents that the files have been calibrated and flat-fielded using \HST{}'s CALWF3 pipeline). 
\texttt{Arctor} examines the header files from the FLT files to extract
\begin{itemize}
    \item The \texttt{EXPEND} and \texttt{EXPSTART} to estimate the observational time as  \texttt{time$_k$} $= \frac{1}{2}\left(  \texttt{EXPSTART}_k +  \texttt{EXPEND}_k\right)$.
    \item \texttt{POSTARG1} and \texttt{POSTARG2}, which represent the spatial location of the initial pointing (before scanning) for both the \texttt{forward} and \texttt{reverse} scans. 
\end{itemize}
Pairs of (\texttt{POSTARG1}, \texttt{POSTARG2}) took exactly two unique pairs of values, which \texttt{Arctor} stores to identify the corresponding indices for \texttt{forward} and \texttt{reverse} scan directions. 
For our \PlanetName{} eclipse observation, we concluded that \texttt{forward} scan trace images began with \texttt{POSTARGS}:[-53.459629, -36.240479]; and \texttt{reverse} scan trace images began with \texttt{POSTARGS}:[-73.743279, -37.577122].
These values are {unique} to each individual observations; but are static values throughout each visit.

\vspace*{5mm}\noindent\textbf{\textit{Cosmic Ray Rejection}}\\
We performed cosmic ray rejection with sigma-outlier estimation over the temporal axis; i.e.~cycling through each pixel, computing the median and standard deviation along the time axis. 
Any values above 5-$\sigma$ from the median were flagged as cosmic rays. 
We then set those pixels to the temporal median.
Although there is an integrated separation of $\sim$344~ppm between the \texttt{forward} and \texttt{reverse} scans in the time series, the median flux variation over time was $\sim$10\% per column,
which is several orders of magnitude greater than the ppm separation between \texttt{forward} and \texttt{reverse} scans.
As such, \texttt{Arctor} performed the cosmic ray rejection over the pixel-by-pixel light curves without considering scan direction.

\vspace*{5mm}\noindent\textbf{\textit{Trace Position and Angle}}

To derive the \textbf{y-center position} of each trace in our images, \texttt{Arctor} fit a 1D Gaussian to all 951 columns in each of the 75 frames, using the {\texttt{Astropy} package\footnote{astropy.org}; \texttt{Astropy} is a standard astronomical functions library \citep{Astropy2018AJ, 2020SciPy-NMeth}}.
We then fit a straight line to the collection of 1D Gaussian \texttt{mean} values for each frame (over the trace columns only), using linear least-squares regression.
Only the \texttt{mean} values for 1D Gaussian fits over the illuminated part of the trace were used in determining the center position and angle (see below).

\begin{figure}
\centering
\includegraphics[width=\linewidth]{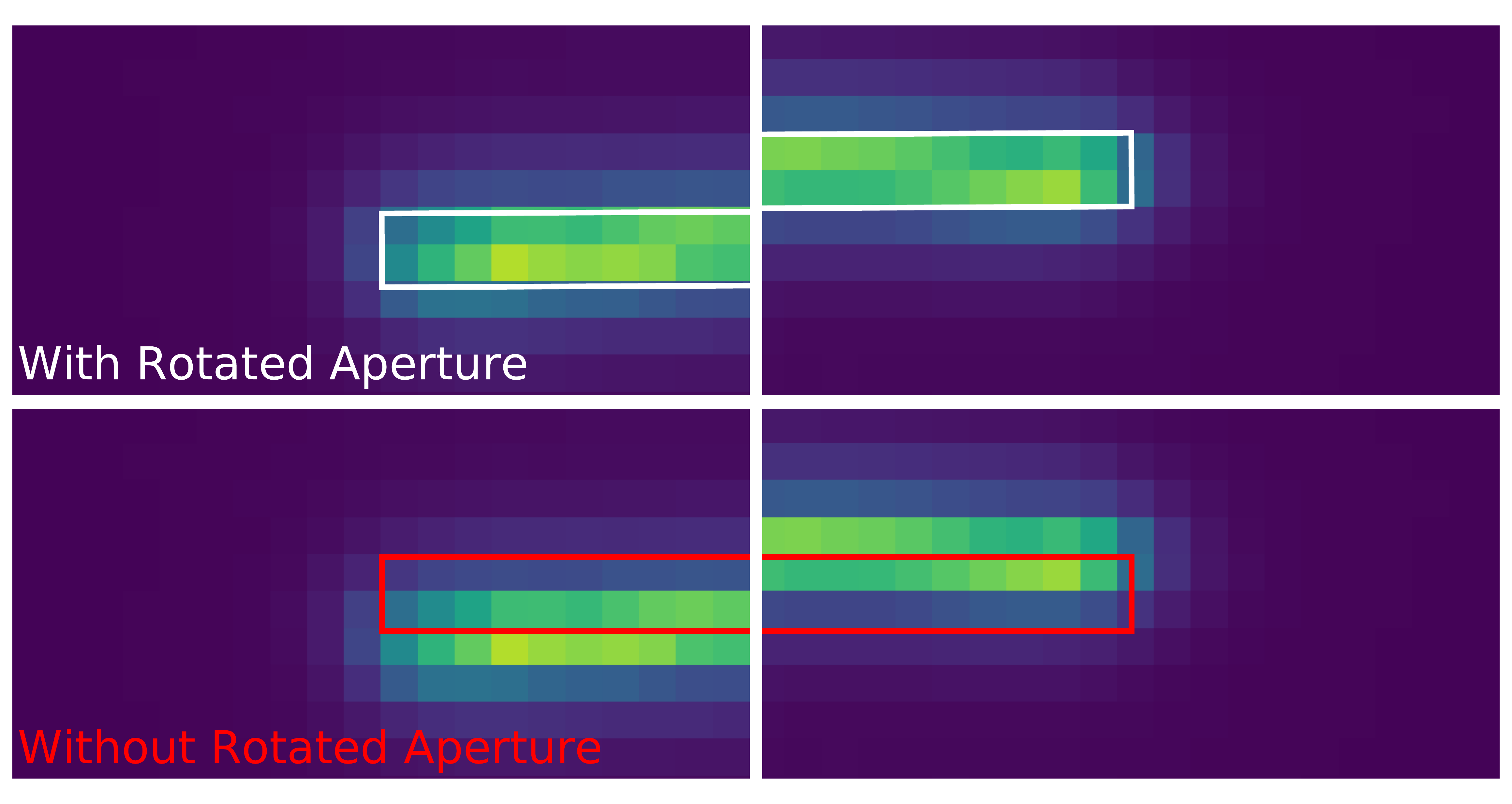}
\caption{The effect of measuring the trace rotation angle. 
The necessity and accuracy of measuring the trace rotation angle for narrow apertures is apparent. 
Although our best \texttt{AICc} \& \texttt{BIC} aperture used a much larger height and width than shown here, the effect of modeling the angle of the trace improved the photometric stability of our light curves by $\sim1$\% (i.e.~5-10~ppm \texttt{SDNR}).}
\label{fig:trace_angles_2D_image}
\end{figure}

{\texttt{Arctor} stored the \texttt{intercept} from this straight line fit as the Y-Center position for each trace; and the slope as the rotation \textbf{angle of the trace}}.
We later used the rotation angle in the photometry and sky background procedures (below) for the \texttt{theta} rotation of the \texttt{rectangular aperture} inside the \texttt{Photutils} package {\citep{Bradley2019Zenodo}}.
\autoref{fig:trace_angles_2D_image} shows the necessity and accuracy of measuring the trace rotation angle for narrow apertures. 
Although our best \texttt{AICc} \& \texttt{BIC} aperture used a much larger height and width than shown here, the effect of modeling the angle of the trace improved the photometric stability of our light curves by $\sim0.5$\%; i.e.~$\sim$10~ppm \texttt{SDNR} over the lightcurve.

\begin{figure}
\centering
\plottwo{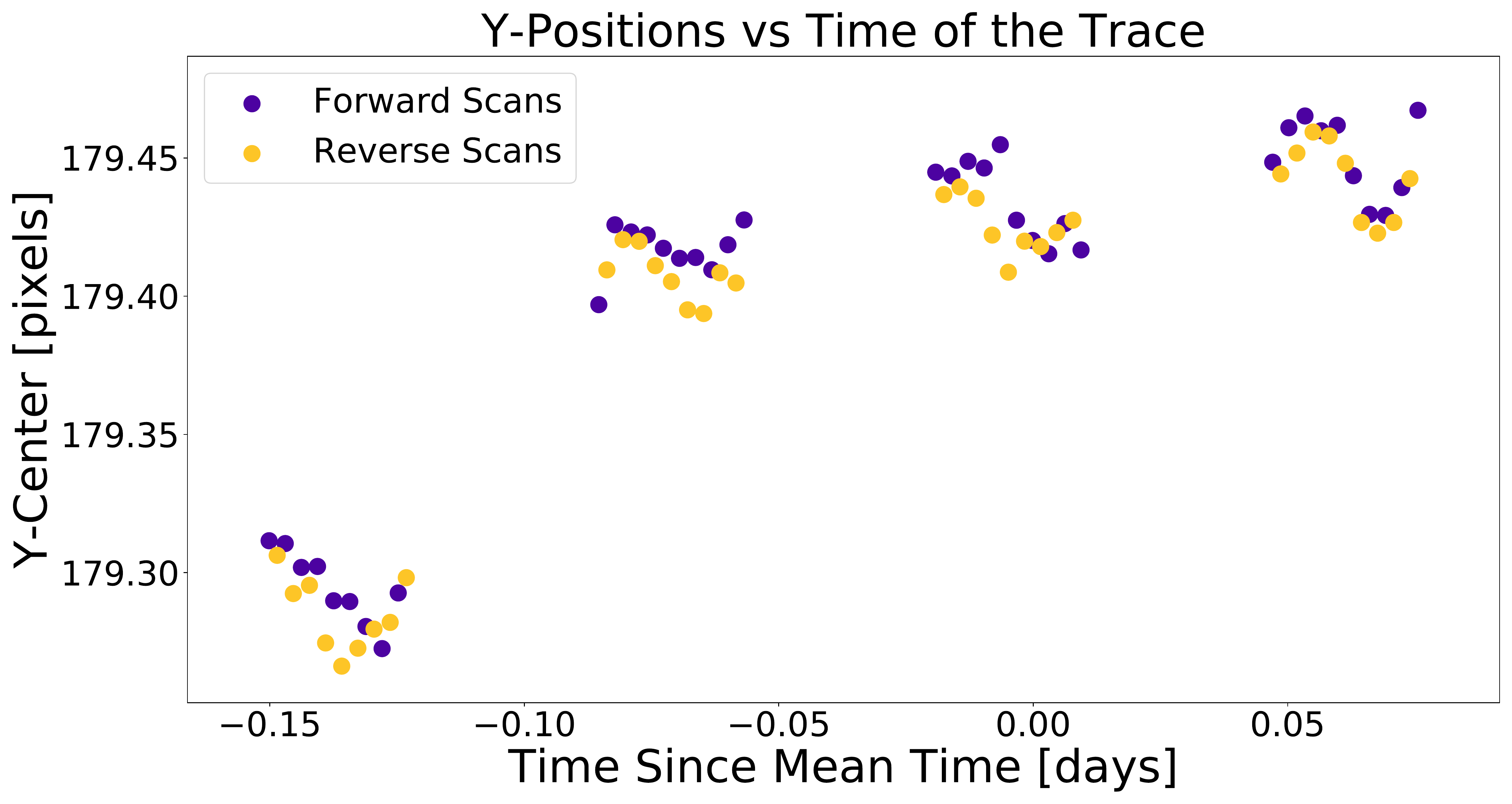}{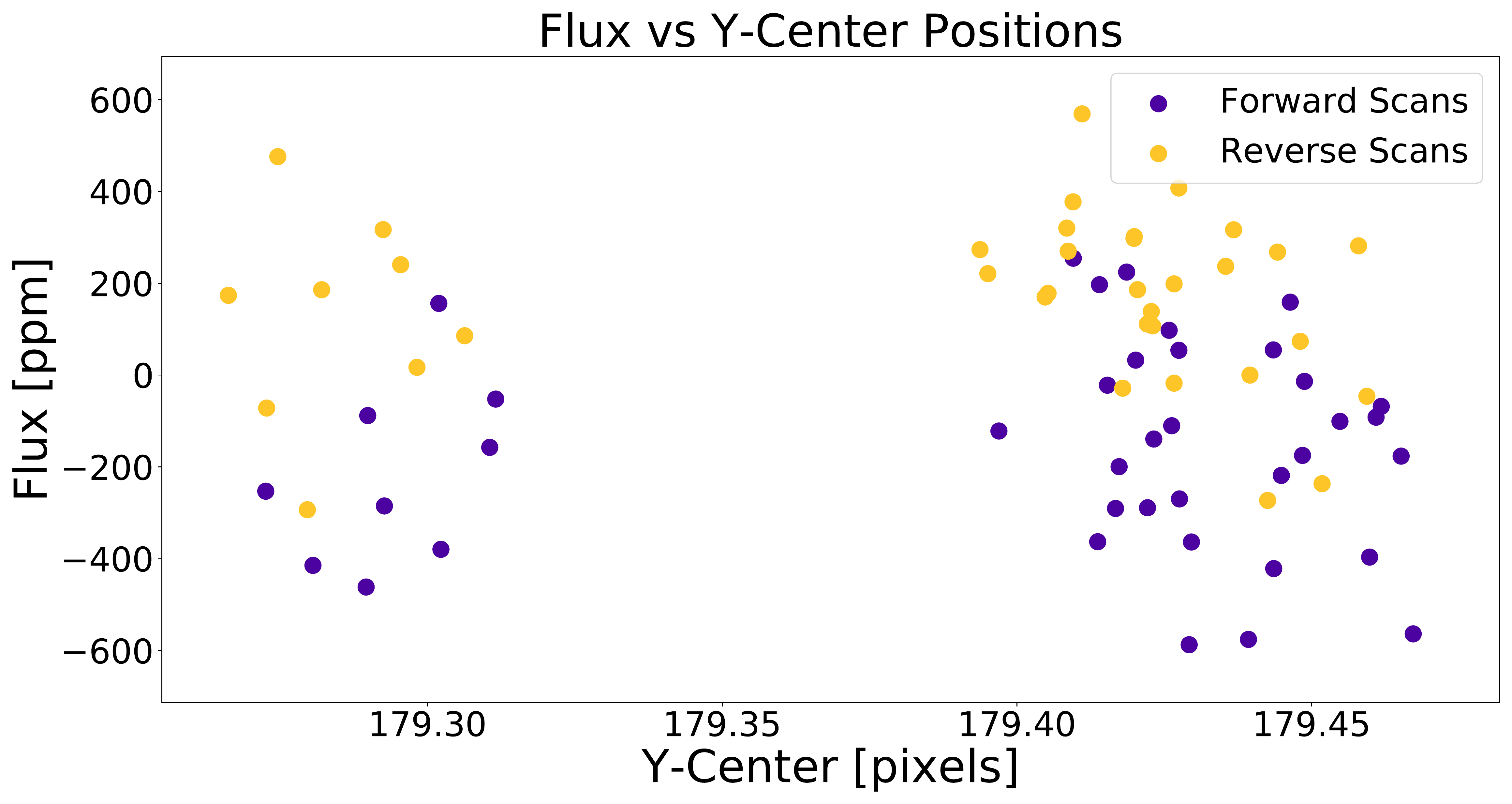}
\caption{{\it [Left]} The Y-center position of our \StarName{} \HSTWFCTUVIS{} observations over 4 \HST{} orbits.
The first orbit sustained Y-center positions significantly different from the three following orbits.
{\it [Right]} Our AICc/BIC model selection process diagnosed a meaningful, linear correlation between the flux and the Y-center positions.
}
\label{fig:y_center_vs_all}
\end{figure}

The y-positions in the first orbit are offset by $\sim$0.1 pixels, or 10$\times$ the inter-orbit scatter of the y-positions -- compared to the three remaining orbits (see \autoref{fig:y_center_vs_all}).
We later derived a meaningful correlation between the y-positions and the flux measured that our \texttt{AICc} \& \texttt{BIC} analysis determined was necessary to select the 'best' model+light~curve pair, implying that including the first orbit in our observations provided improved the SDNR over the full eclipse observation.

\texttt{Arctor} derived the \textbf{x-center positions} by integrating each image in a very narrow subframe that included the trace: a 10 x 951 pixels window.
By integrating down the column within each subframe, the code derived a 1D representation of the trace (integrated flux vs pixel).  
After which, we created a cubic spline to oversample the trace by a factor of 100.
\texttt{Arctor} then measured where the integrated trace reached $>$50\% of its own maximum value on the left and right edges.
It stored the x-center positions as the midpoint between the left and right edges of the 1D trace.
We later determined that this feature vector was critical in reducing correlated noise sources between the \texttt{forward} and \texttt{reverse} scanned images.

\vspace*{5mm}\noindent\textbf{\textit{Investigating Correlations Between Positions and Flux}}

\begin{figure}[t]
\centering
\includegraphics[width=0.85\textwidth,height=0.8\textheight,keepaspectratio]{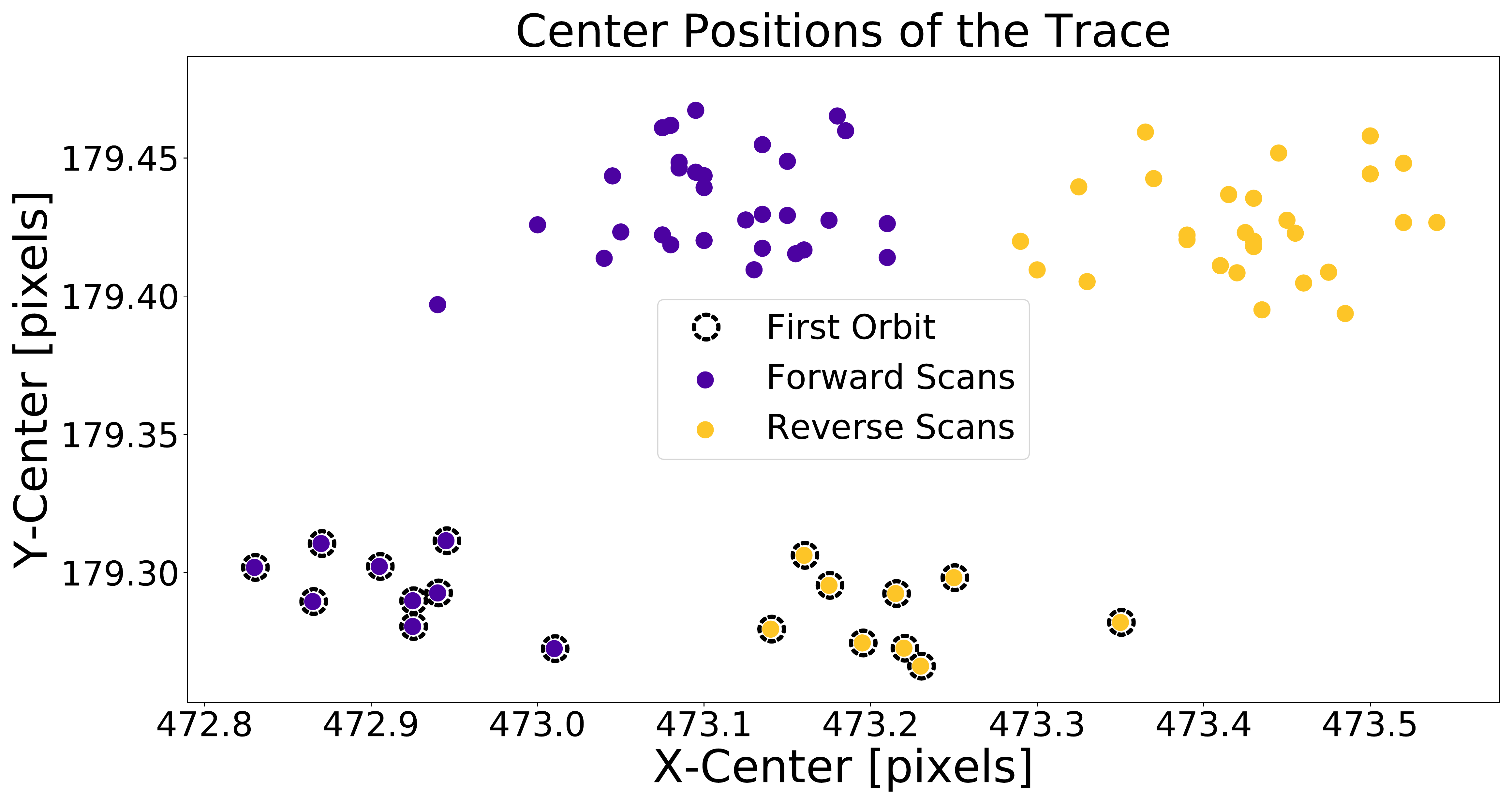}
\caption{The X- and Y-center positions of the trace during all 75 images taken with \HSTWFCTUVIS{} for our \PlanetName{} eclipse observations. 
There is a clear separation of $\sim\frac12$ pixel in x-center position between the forward and reverse scanned images; there is also an small, but significant shift in the y-center positions during the first orbit, as discussed above.
}
\label{fig:ycenter_vs_xcenter}
\end{figure}

After performing a simple integration to estimate the flux (i.e.~Sum(frame - Median(frame))) over time, we discovered a difference in the flux read from \texttt{reverse} scanned images compared the \texttt{forward} scanned images. The separation in median flux measurements between \texttt{reverse} - \texttt{forward} scans is $\sim$344~ppm. 
This could be attributable to differential flat fielding errors encountered by the trace encompassing slightly different pixels during forward and reverse scans (i.e~$<0.5$ pixel difference).
\autoref{fig:ycenter_vs_xcenter} shows the position of all 75 \HSTWFCTUVIS{} images taken during our observations.
There is a clear separation of $\sim\frac{1}{2}$ pixel in x-center position between the forward and reverse scanned images.
There is also an apparent shift in the y-center positions during the first orbit.

The difference in flux could also be related to the lack of shutter, which would cause extra flux to be read by the detector in one of the scan directions.
Specifically, because WFC3 does not have a shutter, there is a slight variation in the exposure time that depends on whether the scan direction is upstream or downstream relative to the detector readout direction.
Examining the trace length vs scan direction (see \autoref{fig:correlation_trace_lengths} below), we noted that \texttt{reverse} scanned images sustained longer traces by $\sim0.2$ pixels, as compared to the \texttt{forward} scaned images.
We accounted for each of these factors in our MAP and MCMC analyses,  using linear models for all five possible parameters (see Section A.1 below).

\vspace*{5mm}\noindent\textbf{\textit{Median Sky Background}}\\
Our purpose here is to measure the integrated background flux inside an rectangular annulus, centered on the trace, and rotated to match the trace. 
We later compare these results with the column-wise sky background estimates (see below).

After measuring the x/y-centers and rotation of the trace, we used a rectangular aperture to include the flux from the scan, while minimizing the background flux included.
The inner aperture was 542 pixels wide by 225 pixels high; the outer aperture was 617 pixels wide by 375 pixels high.
We chose these values to be far outside of the stellar flux (which extends 25 pixels in all directions from the trace) and to avoid edge effects that may occur within $\sim$10 pixels of the edge of the detector.
\autoref{fig:image_and_aperture} shows the exact aperture sizes and relative locations used in measuring both the background sky values and the photometric light curve flux values.


By integrating over the two apertures, then subtracting the medium aperture integration from the larger aperture integration, \texttt{Arctor} generated a mean sky background estimate.
From the \texttt{Photutils} package, there exists a \texttt{rectanguler annulus} function {\citep{Bradley2019Zenodo}}; but it makes assumptions about the ratio between the inner and outer height relative to the inner and outer width that we found did not include enough valuable background real estate.
The result was an average background level of 15.2 e- per pixel.

\vspace*{5mm}\noindent\textbf{\textit{Column-wise Sky Background}}\\
Because our observations use \HST{} scanning mode, there remains the possibility that the background can change as a function of column number (in the scan direction), as a function of either flat-field errors or scattered light that is also being scanned along the image.
As such, \texttt{Arctor} computed both the median and column-wise sky background estimate.

For the column-wise sky background estimation, we used a column-wise aperture for its mask over each image.
We selected the column-wise aperture to include all pixels that are 150 rows above \& below the trace, 
but not including the 10 pixels at the edge, as recommended by the \HST{} scanning mode suggested reduction pipeline.
It stored the 951 median values, taken down the masked columns, as the column-wise sky background vectors, per frame.
We did not rotate the column-wise integration to match the trace angle, in favor of a ``straight down the columns'' median.
The column-wise background estimates averaged closer to 19.5 e- per pixel.
The values included structure ranging from 10 e- at the left edge, 26 e- in the region above/below the trace, and 12 e- on right edge.

After computing our intiial 800 photometry estimates with both the median sky background and the column-wise (1600 photometric estimates combined), we found that both background estimates found equivalent time series light curves raw RMS.
As such, we chose to move forward with the column-wise sky background estimates because they seemed to follow the physical trend of the light on the detector (i.e.~larger in the middle; smaller on the edges). 

\vspace*{5mm}\noindent\textbf{\textit{Estimating the Light Curves: Photometry}}\\
To integrate the flux received from the host star \StarName{} before, during, and after the eclipse, \texttt{Arctor} used the \texttt{rectangular aperture} function to sum over all rows and columns within \texttt{aper-width} and \texttt{aper-height} pixels around the trace (see \autoref{fig:image_and_aperture}).
The apertures were centered on the trace center and rotated to match the rotation of the trace (see \autoref{fig:trace_angles_2D_image}).

\vspace*{5mm}\noindent\textbf{\textit{Coarse Grain Photometry}}\\
We spanned our coarse grain values for \texttt{aper-width} and \texttt{aper-height} between 1-100 and 1-300, respectively -- each encompassing 20 samples; thus the coarse grain \texttt{aper-width} and \texttt{aper-height} spacing was set to 5, and 15 pixels, respectively.
This created 400 light curves that we used to determine the best set of values to center our fine-grain photometry (see below).
Via the standard deviation of raw flux, we computed the scatter along the integrated time series as our metric to select the center and range of our (below) fine-grain \texttt{aper-width} and \texttt{aper-height} values.
While grid searching over both fine grain and coarse grain photometry, we included a range of aperture widths and heights, from 1 pixel up to 400 pixels \textit{outside the trace}.

\vspace*{5mm}\noindent\textbf{\textit{Fine Grain Photometry}}\\
The coarse-grain span of photometric light curves revealed that \texttt{aper-width} and \texttt{aper-height} values of 21 $\times$ 51, respectively, resulted in the minimum scatter (std-dev) of $206$~ppm.
As a result, we recomputed 400 new light curves with \texttt{aper-width} and \texttt{aper-height} values spanning from 11-31 and 41-61, respectively -- encompassing 20 samples in each parameter; thus the fine grain \texttt{aper-width} and \texttt{aper-height} spacing was set to 1 pixels.
These integrations also found the 'best' light curve scatter of $206$~ppm; but
with  \texttt{aper-width} and \texttt{aper-height} values of 13 $\times$ 45.
This is the light curve that we show in Figures~\ref{fig:raw_flux}~and~\ref{fig:best_AIC_flux_model}.

\vspace*{5mm}\noindent\textbf{\textit{Maximum A Posteriori (MAP)}}\\
Using the \texttt{exoplanet}\footnote{https://github.com/dfm/exoplanet} package, we fit 32 models to each of the 400 light curves, culminating in 12800 model fits. The 32 models were $2^5$ toggles for every possible combination of whether to simultaneously fit a linear trend to any of our 5 systematic features: \texttt{Fwd/Rev Indices}; \texttt{X-Centers} (\autoref{fig:correlation_xcenter}); \texttt{Y-Centers} (\autoref{fig:correlation_ycenter}); \texttt{Trace Angles} (\autoref{fig:correlation_trace_angles}); and \texttt{Trace Lengths} (\autoref{fig:correlation_trace_lengths}).
We analyzed our exracted dataset of AICc/BIC results per MAP fit to select the \textit{best} model, which minimized both AICc and BIC.
Note that a linear trend was fit to the \texttt{time} feature in all cases; this temporal variation can be attributed to long term variation in the sensitivity or flat field error; or the stellar variability, if significant enough.

In \autoref{appendix:map_aicc} we show the result of all 12800 models, compared across \texttt{SDNR}, $\chi^2$, \texttt{AICc}, and \texttt{BIC}.  In the \texttt{AICc} \& \texttt{BIC} case, the global picture and `best' model selected were identical; i.e.~a `13x45' sized rectangular aperture. Moreover, linear fits to \texttt{x-center}, \texttt{y-center}, and \texttt{trace-length} resulted in the minimum \texttt{AICc} and \texttt{BIC} light curve, with \texttt{SDNR}~= ~72~ppm; see \autoref{fig:best_AIC_flux_model}. The cohesion between complementary information criteria strongly implies that our model selection process is valid and sound.

\vspace*{5mm}\noindent\textbf{\textit{Markov Chain Monte Carlo Posterior Estimation}}\\
{To extract the greatest information content, we used a collection of 16 Hamiltonian Monte Carlo (HMC) estimates using linear and non-linear models.
We sampled 3000 \textit{tune} steps and 3000 \textit{draw} steps for all 16 chains and each of the 25 light curves examined with this HMC; increasing the tune/draw counts did not change the results.
We used a log-normal prior for the \textit{Mean Offset}, \textit{Eclipse Depth} and a wide Normal prior for the four slope parameters that fit as well: slopes over the \textit{Time}, \textit{X-Center}, \textit{Y-Center}, \textit{Trace-Length} features.
We also tested a log-uniform and uniform prior over the \textit{Eclipse Depth}.
After sampling 16 chains per each of the 25 light curves, we used the autocorrelation time scale and Gelman-Rubin Test, as provided by the \texttt{exoplanet} package \cite{DFM2019ascl}.
This package includes an analytic transit model (\texttt{BATMAN}; \citealp{Kreidberg2015PASP}) and a spherical harmonic phase curve model (\textit{STARRY}; \citealp{Luger2019AJ}).}


\autoref{fig:corner_plot_best_AIC} shows the corner plot (i.e.~2D MCMC  correlation plots) from our best \texttt{AICc} \& \texttt{BIC} selected model + light curve pair. 
Our best model fit for the \texttt{Mean} (i.e.~global flux offset), \texttt{Edepth} (i.e.~the depth of the Eclipse), \texttt{Slope} (in \texttt{time}), \texttt{Slope Xcenter}, \texttt{Slope Ycenter}, \texttt{Slope Trace Length}; the slopes were generated from the simultaneous linear fits to the corresponding features, with a single, unified `intercept' represented by the \texttt{Mean}.
The results clearly show that the slopes in each of the three feature spaces were significantly detected and contributed to the quality of the final SDNR; but that the eclipse depth was \textbf{not} detected, with a $3\sigma$ upper limit of 67~ppm.

Although we do not show this here, we also fit to the eclipse depth with a uniform and log-uniform prior. The log-uniform prior also derived a non-detection, with a peak around $\log10$(eclipse depth) $\sim-20$; i.e.~an eclipse depth $\sim10^{-20}$. Unfortunately, the log-uniform prior appeared to be more strongly influenced by the exact bounds of the prior; i.e.~the MAP would vary from $\log10$(eclipse depth) of -100 to -10, depending on the selected prior boundaries. In contrast, the uniform prior for the eclipse depth provided consistent results across all of our trials and selections for the prior boundaries.

\subsection{\HSTWFCTUVIS{} Scanning Mode Correlated Noise}
\begin{figure}
\centering
\includegraphics[width=\linewidth]{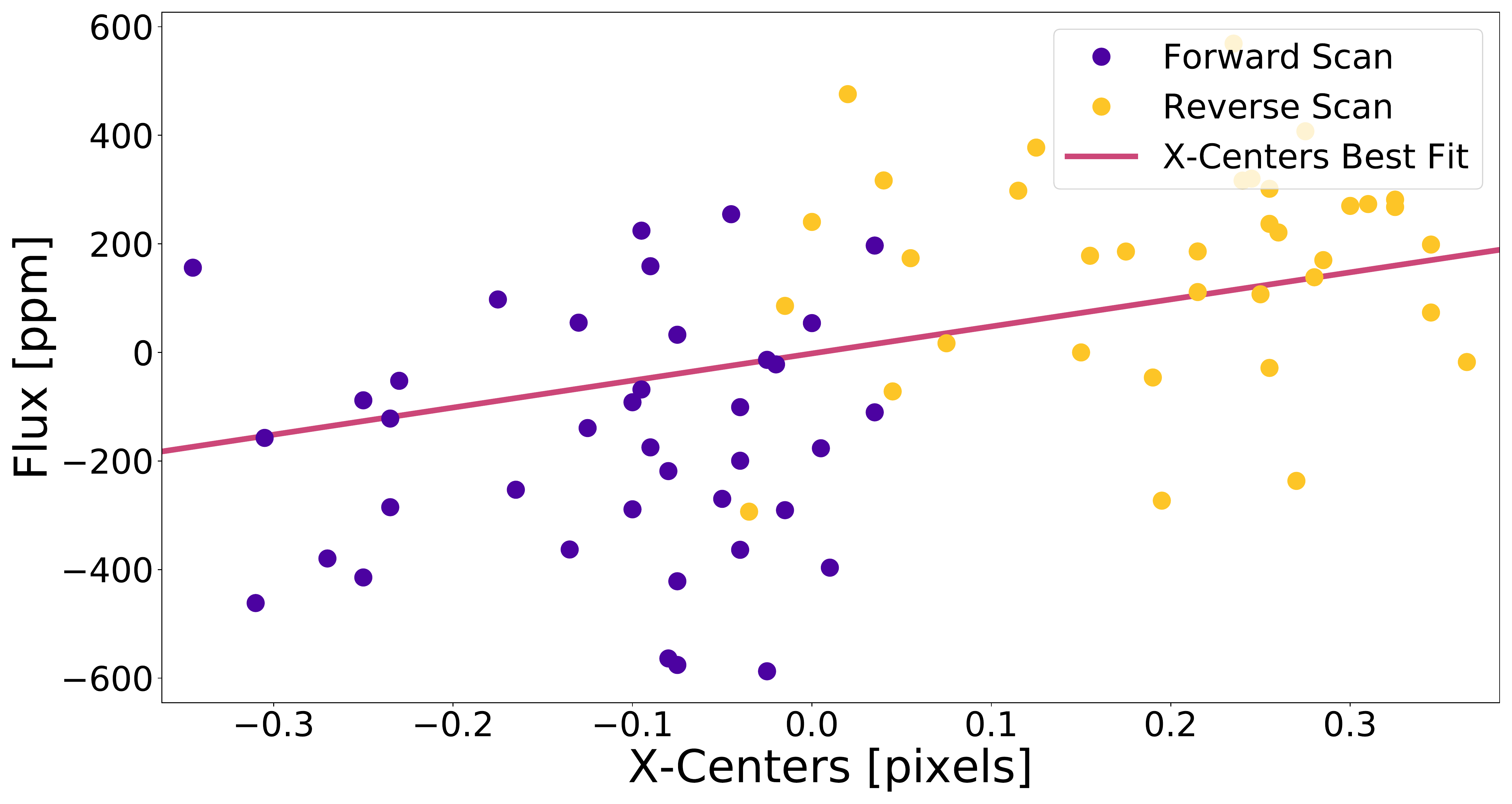}
\caption{2D linear correlation between Time + X-Center positions and Normalized Flux for our (\texttt{AICc}) `best' light curve. The slope values are relative to the standard deviation of the flux and pixel coordinates (over the full observation). As before, we coded the {forward scan values in violet} and  {reverse scan values in orange}.
}
\label{fig:correlation_xcenter}
\end{figure}

\begin{figure}
\centering
\includegraphics[width=\linewidth]{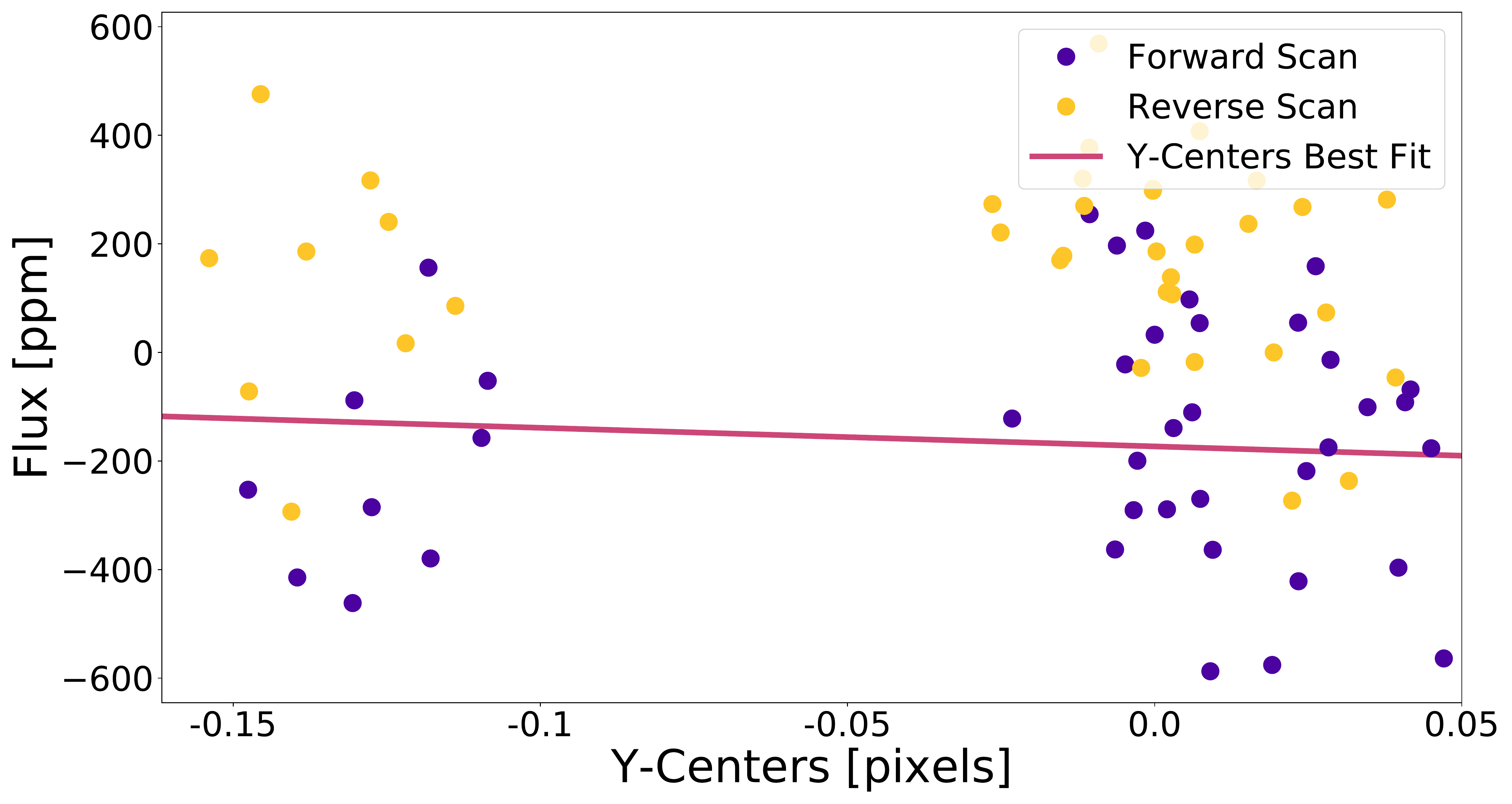}
\caption{2D linear correlation between Time + Y-Center positions and Normalized Flux for our (\texttt{AICc}) `best' light curve. The slope values are relative to the standard deviation of the flux and pixel coordinates (over the full observation). As before, we coded the {forward scan values in violet} and  {reverse scan values in orange}.
}
\label{fig:correlation_ycenter}
\end{figure}

\begin{figure}
\centering
\includegraphics[width=\linewidth]{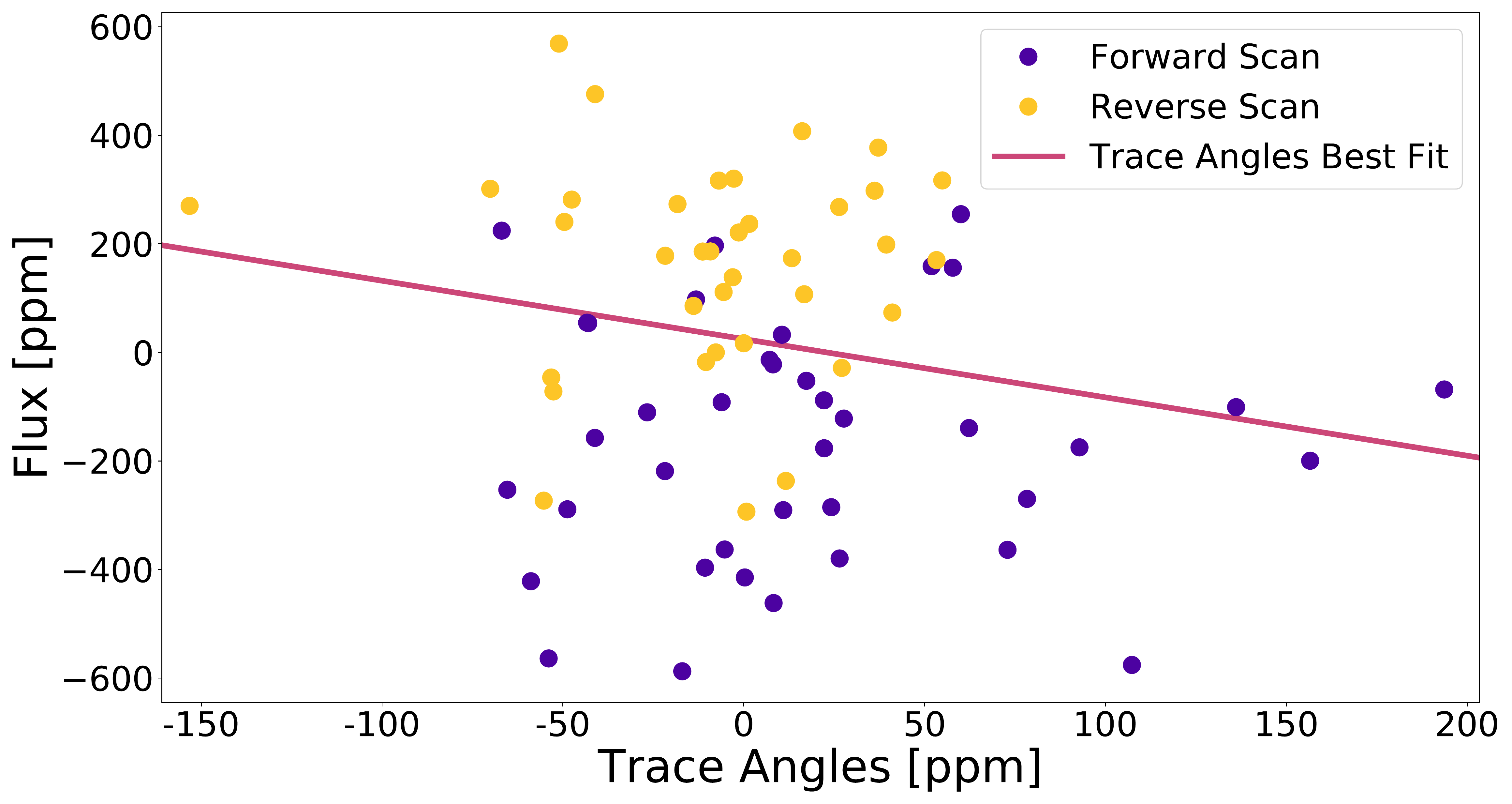}
\caption{2D linear correlation between Time + Trace Angles and Normalized Flux for our (\texttt{AICc}) `best' light curve. The slope values are relative to the standard deviation of the flux and pixel coordinates (over the full observation). As before, we coded the {forward scan values in violet} and  {reverse scan values in orange}.
}
\label{fig:correlation_trace_angles}
\end{figure}

\begin{figure}
\centering
\includegraphics[width=\linewidth]{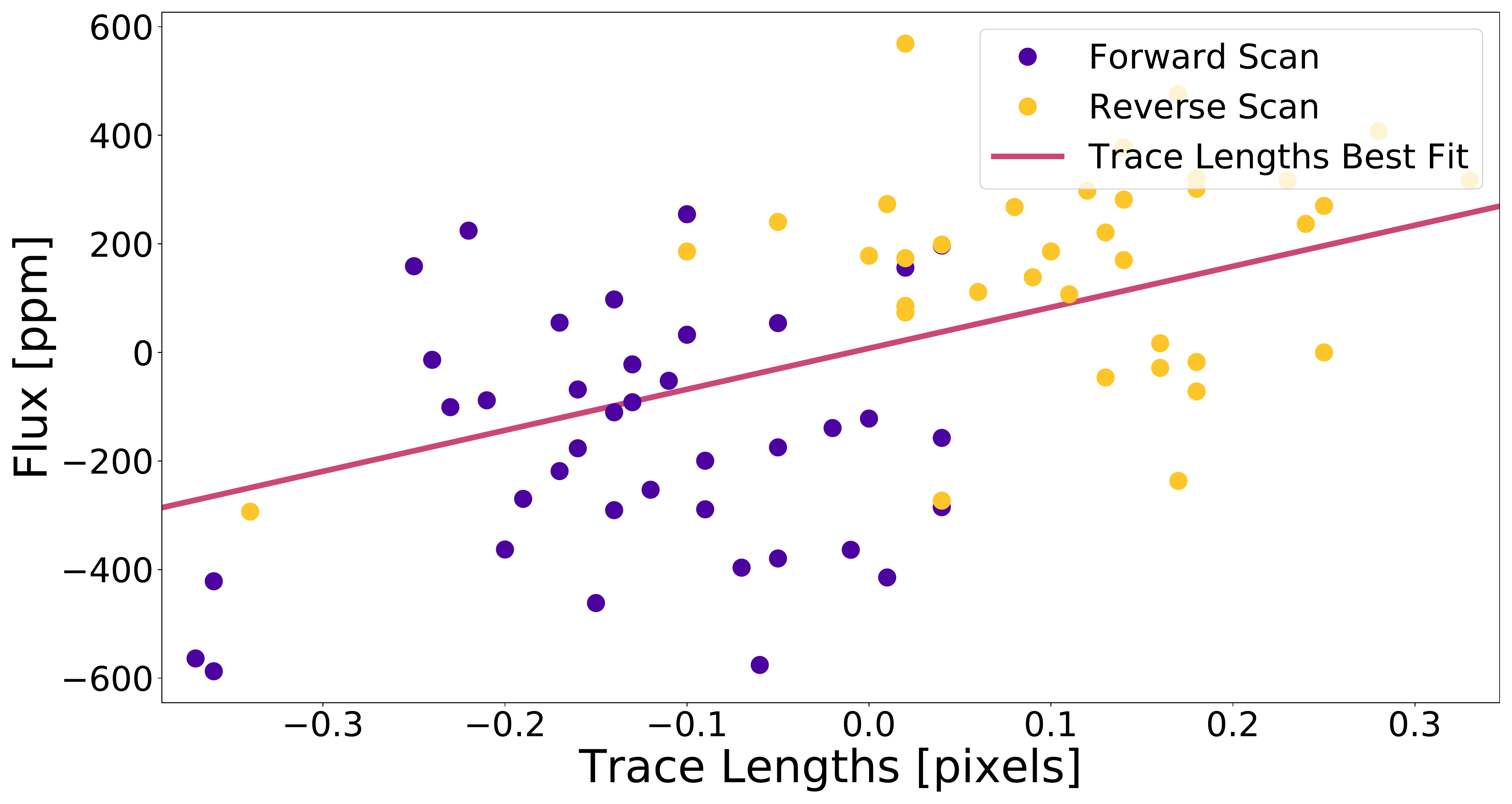}
\caption{2D linear correlation between Time + Trace Lengths and Normalized Flux for our (\texttt{AICc}) `best' light curve. The slope values are relative to the standard deviation of the flux and pixel coordinates (over the full observation). As before, we coded the {forward scan values in violet} and  {reverse scan values in orange}.
}
\label{fig:correlation_trace_lengths}
\end{figure}

In order to diagnose any plausible connections between the measured flux and non-astrophysical noise sources, we examined a suite of linear trends between the flux measurements and \texttt{Time Values}, \texttt{X-Center Positions}, \texttt{Y-Center Positions}, \texttt{Trace Angles}, \texttt{Trace Lengths}; as seen in Figures~\ref{fig:correlation_xcenter},~\ref{fig:correlation_ycenter},~\ref{fig:correlation_trace_angles},~\ref{fig:correlation_trace_lengths}, respectively.
These figures show initial estimates for 2D linear correlations between normalized, median subtracted, time feature + \textit{other} feature onto the flux measurements, for each of the four features. 
That is, we fit a plane to the features, including  \texttt{time} in all four fits, while alternating the other four features listed above.

We also examined the effect of fitting two distinct \texttt{means} for flux values corresponding to the \texttt{FWD \& REV Indices}; i.e.~a \texttt{mean$_{rev}$} fit to the orange points in \autoref{fig:raw_flux} and a \texttt{mean$_{fwd}$} fit to the violet points in \autoref{fig:raw_flux}. 
When considering separated \texttt{mean} flux values, we fit for the one set of eclipse parameters and slope values over each systematic feature.
These fits were included the $2^5$ combinations, mentioned above, in our \texttt{AICc} \& \texttt{BIC} model selection process.

The most significant linear trend that we detected from only a time~+~feature vs flux, resulted from bilinear fitting with \texttt{X-Center Position} and \texttt{Time Value} features, compared to normalized flux estimates.
Note that Planar2D is a plane model (i.e.~bi-linear:  \text{Flux} $=$ a $\times$  \text{Time} + b $\times$ \text{Other}), for use with fitting routines.

Figures~\ref{fig:correlation_xcenter},~\ref{fig:correlation_ycenter},~\ref{fig:correlation_trace_angles},~\ref{fig:correlation_trace_lengths} only show our initial estimates from our fits. 
Our final \texttt{AICc} \& \texttt{BIC} and parameter estimates were generated using the combination of Maximum A Posteriori (MAP) and Bayesian Credible Regions (BCR) analysis; both sourced from  the \texttt{exoplanet} package (XO)\citep{DFM2019ascl}.

Using the XO package, we were able to test 12800 model + aperture configurations; i.e.~32 models with 400 rectangular aperture estimates of the light curve for various aperture widths and heights.
We provide a descriptive visualization of these 12800 model results over our information criteria based model selection techniques in \autoref{appendix:map_aicc}.
The result of this analysis revealed that the `best fit model' resulted in a minimum \texttt{AICc}~=~148 and \texttt{BIC}~=~146, with $\Delta$\texttt{AICc}~=~$\Delta$\texttt{BIC}~=~2, which is considered a significant decrease \citep{Schwarz1978, Schoniger2014AICBIC}.

\clearpage
\section{Comparative Maximum A Posteriori + AICc Analysis over Linear Correlations}
\label{appendix:map_aicc}
\subsection{Linear Models}
To select between our 12800 model+light curve pairs (i.e. parameters + hyper-parameters), we computed the AICc and BIC for each of our options. 
We then chose the option with the minimum AICc and BIC. This model resulted from the same model option, which was not guaranteed \textit{a priori}. 
Here we describe our process and the underlying considerations when using AICc and BIC for model selection.

The AICc is the \textit{corrected} Akaike Information Criterion; and the BIC is the Bayesian Information Criterion. 
Both information criteria are sufficient metrics for model selection when comparing analytic, quasi-linear models.
Both IC penalizes the log-likelihood function ($\log{L}$) by the number of parameters ($k$), relative to the number of samples ($n$); i.e.
\begin{equation*}
    BIC = -2 \log{L} + k \log{n}
    \label{eqn:BIC}
\end{equation*}
\begin{equation*}
    AICc = -2 \log{L} + k + \frac{2k\left(k + 1\right)}{n - k - 1}
    \label{eqn:AICc}
\end{equation*}

Each equation is an approximation to the Bayesian evidence under very specific linear and/or Gaussian conditions \citep{Schoniger2014AICBIC}. 
The AICc is \textit{corrected} for usage with a finite number of data points, i.e.~``small'' data sets.
These criteria provide a more robust estimate for the "goodness of fit" than the SDNR (standard deviation of normalized residuals) or $\chi^2$ ("chi-squared"); the latter being a simplified representation of the log-likelihood ($\log{L}$ above) under the assumption that the posterior approximates a Gaussian normal distribution.

We used the AICc and BIC to select which of the 32 models (i.e.~\textit{parameters}) and 400 light curves (i.e.~\textit{hyper-parameters}) provided the \textit{best} representation of the data.
The \textit{best} model + light curve pair achieved the smallest AICc or BIC by at least a $\Delta$BIC/AICc of 2 \citep{Schwarz1978, Schoniger2014AICBIC}.

Each of our models included \textit{up to} five linear parameters + the eclipse depth; i.e.
\begin{table}[h]
\begin{center}




\begin{tabular}{lc}
               & Number of\\
Model Features & Parameters \\
\hline
Time + Eclipse + Offset & 3\\
Time + Eclipse + Offset + XCenters & 4\\
Time + Eclipse + Offset + YCenters & 4\\
Time + Eclipse + Offset + TraceAngles & 4\\
Time + Eclipse + Offset + TraceLengths & 4\\
Time + Eclipse + Offset + XCenters + YCenters & 5\\
Time + Eclipse + Offset + XCenters + TraceAngles & 5\\
Time + Eclipse + Offset + XCenters + TraceLengths & 5\\
Time + Eclipse + Offset + YCenters + TraceAngles & 5\\
Time + Eclipse + Offset + YCenters + TraceLengths & 5\\
Time + Eclipse + Offset + TraceAngles + TraceLengths & 5\\
Time + Eclipse + Offset + XCenters + YCenters + TraceAngles & 6 \\
Time + Eclipse + Offset + XCenters + YCenters + TraceLengths & 6 \\
Time + Eclipse + Offset + XCenters + TraceAngles + TraceLengths & 6 \\
Time + Eclipse + Offset + YCenters + TraceAngles + TraceLengths & 6 \\
Time + Eclipse + Offset + XCenters + YCenters + TraceAngles + TraceLengths  & 7\\
\end{tabular}
\caption{
    The set of 16 linear models tested for our analysis. 
    We used all 32 possible combinations of these features; and only in with linear model; see equation below.
    Time + Eclipse + Offset was our most simple model.
    Moreover, each of these models was also fit with or without separating the dataset into forward and reverse scanned -- independent, but linked -- sub-sets.
    This resulted in 16 models without separating into forward and reverse scans + 16 models with separating into forward and reverse scans.
\label{tbl:models}}
\end{center}
\end{table}
\begin{equation*}
    \text{Flux}=  a\cdot \text{Time} + 
                  b\cdot \text{XCenter} + 
                  c\cdot \text{YCenter} + 
                  d\cdot \text{TranceLength} + 
                  e\cdot \text{TranceAngle} + 
                  \text{EclipseModel} + 
                  \text{Offset}
\end{equation*}

\begin{figure}
\centering
\includegraphics[width=\textheight,angle=90,origin=c]{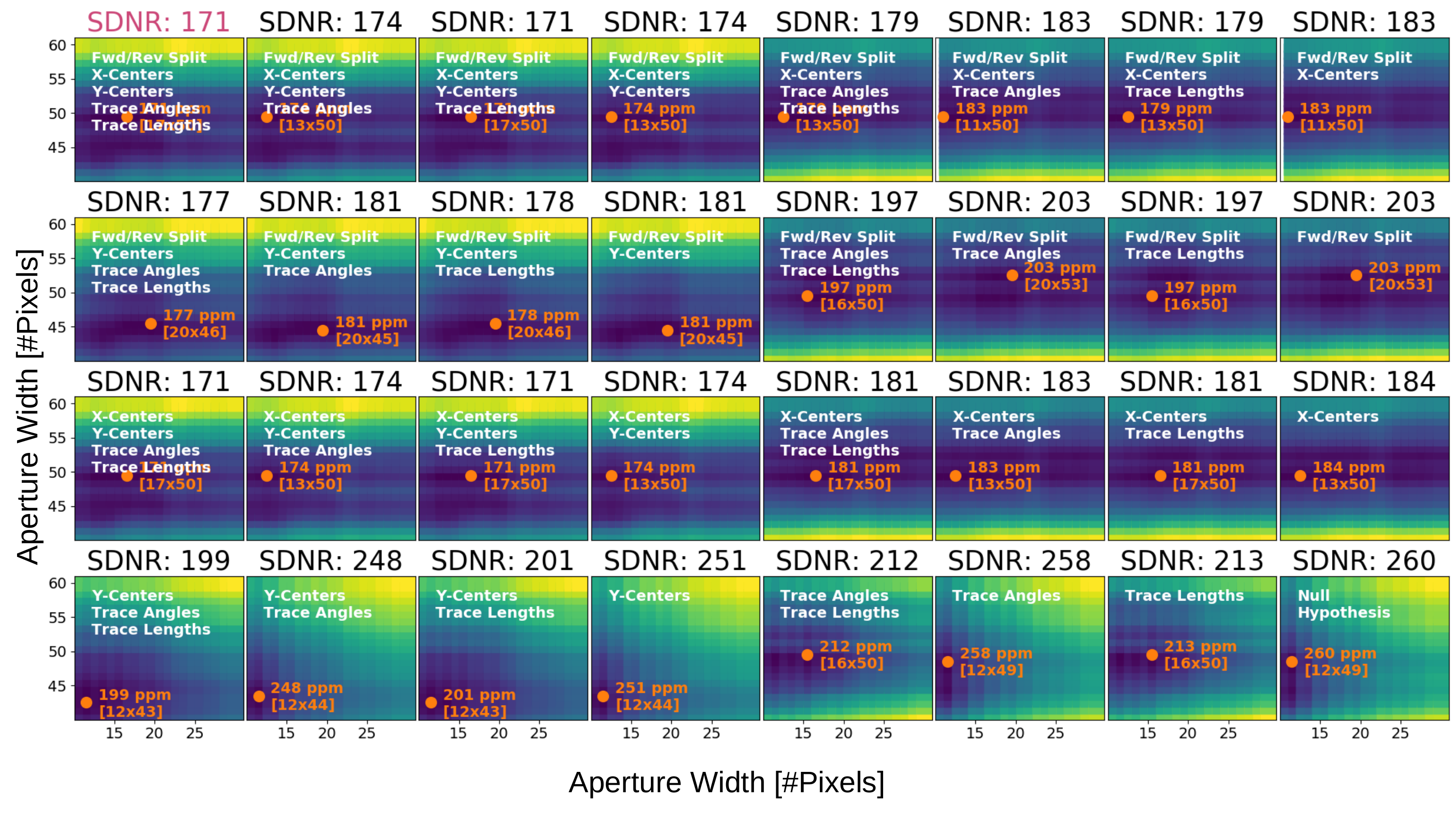}
\caption{12800 MAP results compared to SDNR.
The SDNR will almost always choose the model with the most number of parameters.
This can be seen by the minimum SDNR achieved in the upper left panel: SDNR=171.
We use AICc and BIC to avoid falling into models that do not extrapolate well.}
\label{fig:sdnr_vs_map}
\end{figure}

\begin{figure}
\centering
\includegraphics[width=\textheight,angle=90,origin=c]{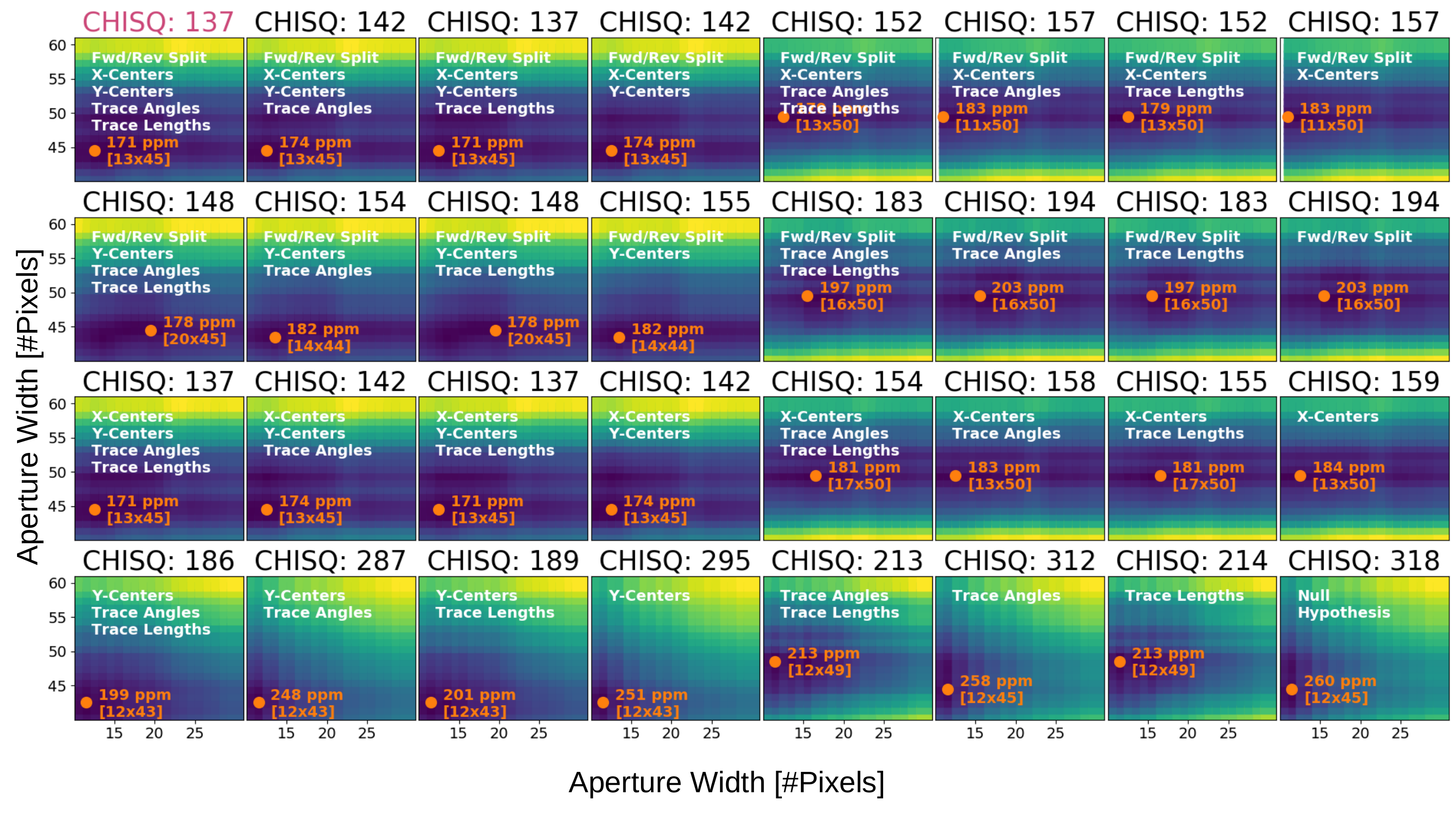}
\caption{12800 MAP results compared to $\chi^2$.
Much like SDNR, $\chi^2$ will almost always choose the model with the most number of parameters.
This can be seen by the minimum $\chi^2$ achieved in the upper left panel: $\chi^2$=137.
We use AICc and BIC to avoid falling into models that do not extrapolate well.}
\label{fig:chisq_vs_map}
\end{figure}

\begin{figure}
\centering
\includegraphics[width=\textheight,angle=90,origin=c]{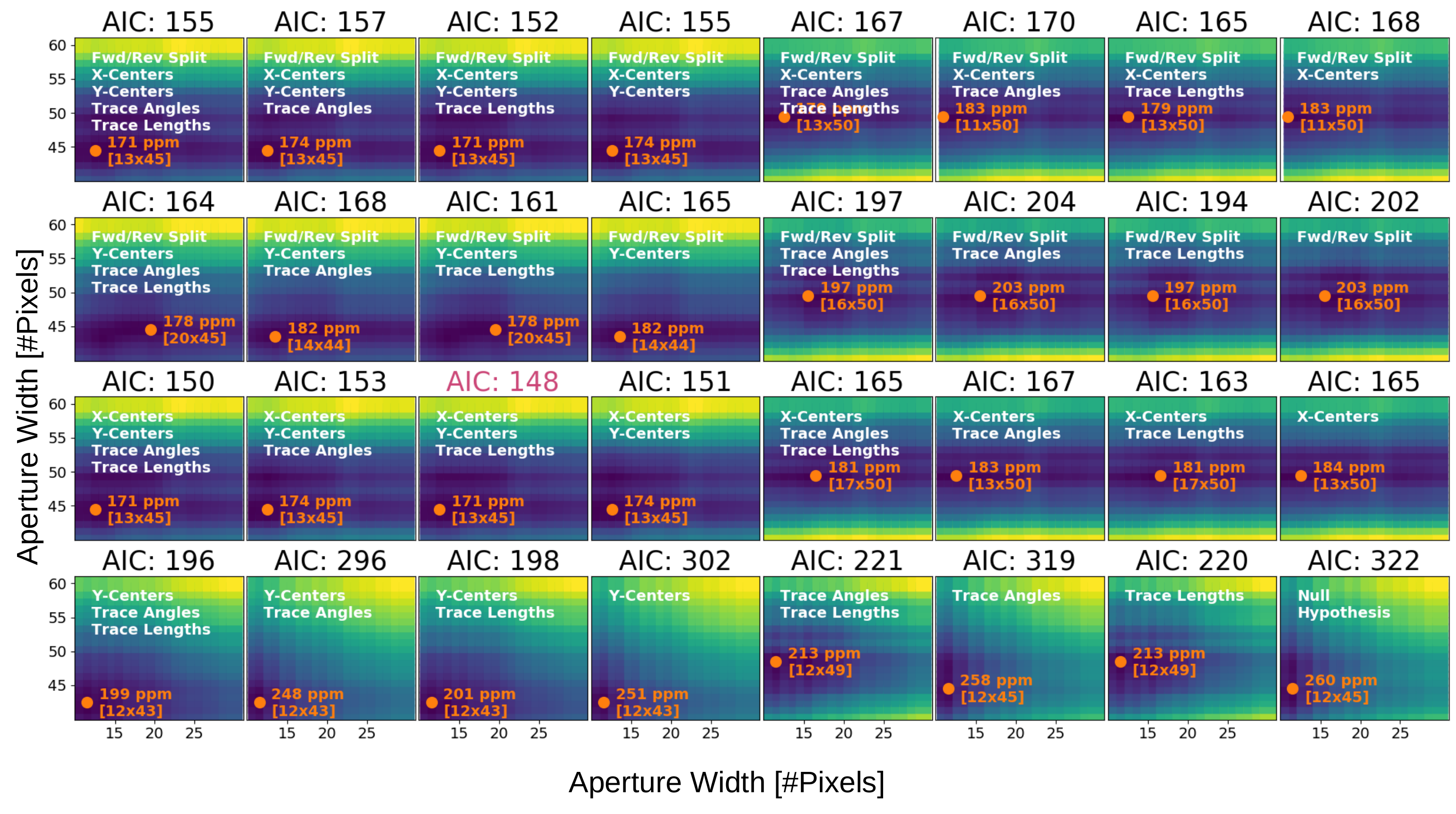}
\caption{12800 MAP results compared to AICc.
Unlike like SDNR or $\chi^2$, the AICc will only choose the model with the most number of parameters \textit{only} if it also minimizes the complexity of the dataset.
This can be seen by the minimum AIC achieved in the (row,col) = (3, 3) [starting from upper left]; AIC=148.}

\label{fig:aic_vs_map}
\end{figure}

\begin{figure}
\centering
\includegraphics[width=\textheight,angle=90,origin=c]{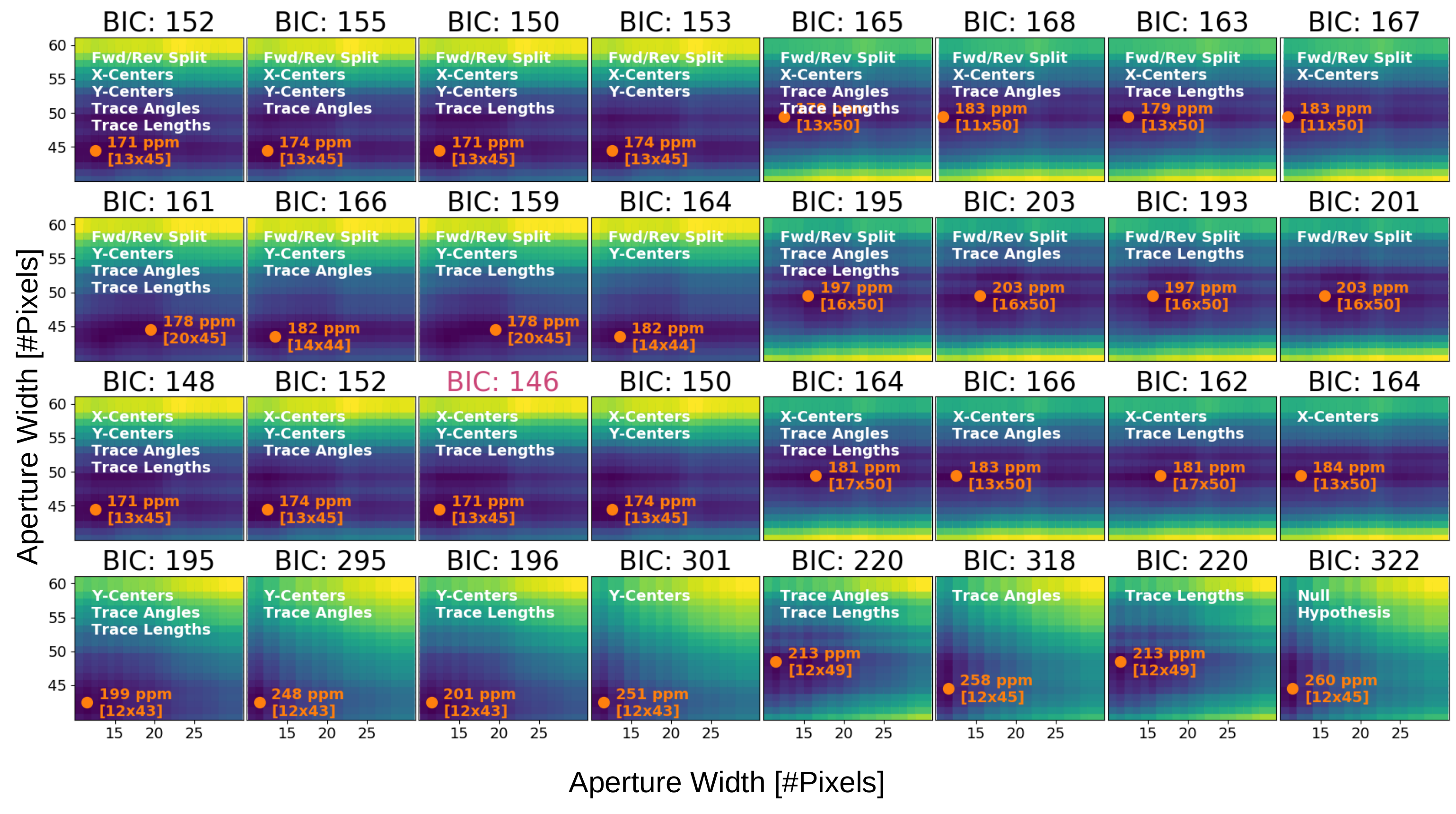}
\caption{12800 MAP results compared to BIC.
Unlike like SDNR or $\chi^2$, the BIC will only choose the model with the most number of parameters \textit{only} if it also minimizes the complexity of the dataset -- the same as AICc.
This can be seen by the minimum BIC achieved in the (row,col) = (3, 3) [starting from upper left]; BIC=146.}
\label{fig:bic_vs_map}
\end{figure}

The \textit{best} AICc and BIC result was achieved using a linear fit to four features: Time, X-Center positions, Y-Center positions, and Trace Lengths.
The follow ing figures show the distribution of AIC, BIC, SDNR, and $\chi^2$ over each of our 12800 models. The important factors to see are that the minimum SDNR (\autoref{fig:sdnr_vs_map}) and $\chi^2$ (\autoref{fig:chisq_vs_map}) are always achieved by the model with the \textbf{most} parameters (i.e. upper left panel) to achieve the least RMS scatter; while the AICc (\autoref{fig:aic_vs_map}) and BIC (\autoref{fig:bic_vs_map}) select the model with a balance between the number of parameters and the complexity of the model; i.e. with an AICc=148 and a BIC=146. [Note that those AICc and BIC cannot be compared directly to each other.]

\subsection{Auto-Correlated Noise}
\label{gaussian_processes}
After selecting the model configuration that minimized the \texttt{AICc} and \texttt{BIC}, we further investigated the possible presence of auto-correlated noise, i.e.~``power-law noise'' with PSD$(\omega)\sim \omega^{-\beta}$ \citep{Pont2006,Cubillos2017AJ}. {We attempted several power-law noise procedures (see below), and included them in the Arctor package; but only the Gaussian Process (GP) analysis produced reproducible MCMC results \citep{DFM2017AJ, DFM2019ascl}.}

\citet{Carter2009ApJ_wavelets} derived a wavelet analysis technique for transiting exoplanet observations with the Spitzer Space Telescope that introduced the terminology $\sigma_r$ to represent the amplitude of residual, power-law noise; and $\sigma_w$ to represent the underlying white noise in the residuals.
That algorithm specifically mitigates `pink noise` (i.e.~PSD$(\omega)\sim \omega^{-1}$); and requires evenly spaced data, with high cadence, and many samples.
\citet{Carter2009ApJ_wavelets, Cubillos2017AJ} discusses several other techniques to mitigate various `power-law noise' effects; they also focused on high cadence, transit observations with the Spitzer Space Telescope. 
Because of HST's large duty cycle and gaps for Earth's eclipse, our data does not satisfy the underlying assumptions that would allow these techniques to be effective.
As such, we chose to implement a Gaussian Process (GP) regression technique that is designed to identify non-periodic, auto-correlated noise signatures in residual data \citep{gibson12a, gibson12b, DFM2017AJ}. 
More directly, we used the implementation of \texttt{celerite} that was built into \texttt{exoplanet}\footnote{https://github.com/exoplanet-dev/exoplanet} \citep{DFM2017AJ}.

Using a simple harmonic oscillator kernel with the \texttt{celerite} package\footnote{https://github.com/dfm/celerite}, we modeled the auto-correlated noise with a covariance matrix kernel shape:
\begin{equation*}
    k(\tau) = S_o\omega_o\exp^{-\frac{\omega_o\tau}{\sqrt{2}}} \cos{\left(\frac{\omega_o\tau}{\sqrt{2}} - \frac\pi4\right)}
\end{equation*}
which obeys the power-law relationship
\begin{equation*}
    S(\omega) = \sqrt{\frac2\pi}\frac{S_o\omega_o^4}{\omega^4+\omega_o^4} \sim \omega^{-4} \text{ for large } \omega
\end{equation*}

We represent $\sigma_r = S_o \omega_o^4$ as the the amplitude of the power-law noise.
We also use an error scaling term, $\sigma_w$, along the diagonal of the covariance matrix: $\sigma_{\text{data}}^2 + \sigma_w^2$\footnote{https://github.com/exoplanet-dev/exoplanet/issues/76}.
Furthermore, we sampled the GP kernel hyper-parameters from wide log-normal priors: $\log{\left(\sigma_r\right)} \sim \mathcal{N}\left(0, 15\right)$ and $\log{\left(\sigma_w\right)} \sim \mathcal{N}\left(0, 15\right)$, which created log-normal distributions from (0, $\infty$) centered at unity.

\begin{figure}
\centering
\includegraphics[width=\textwidth]{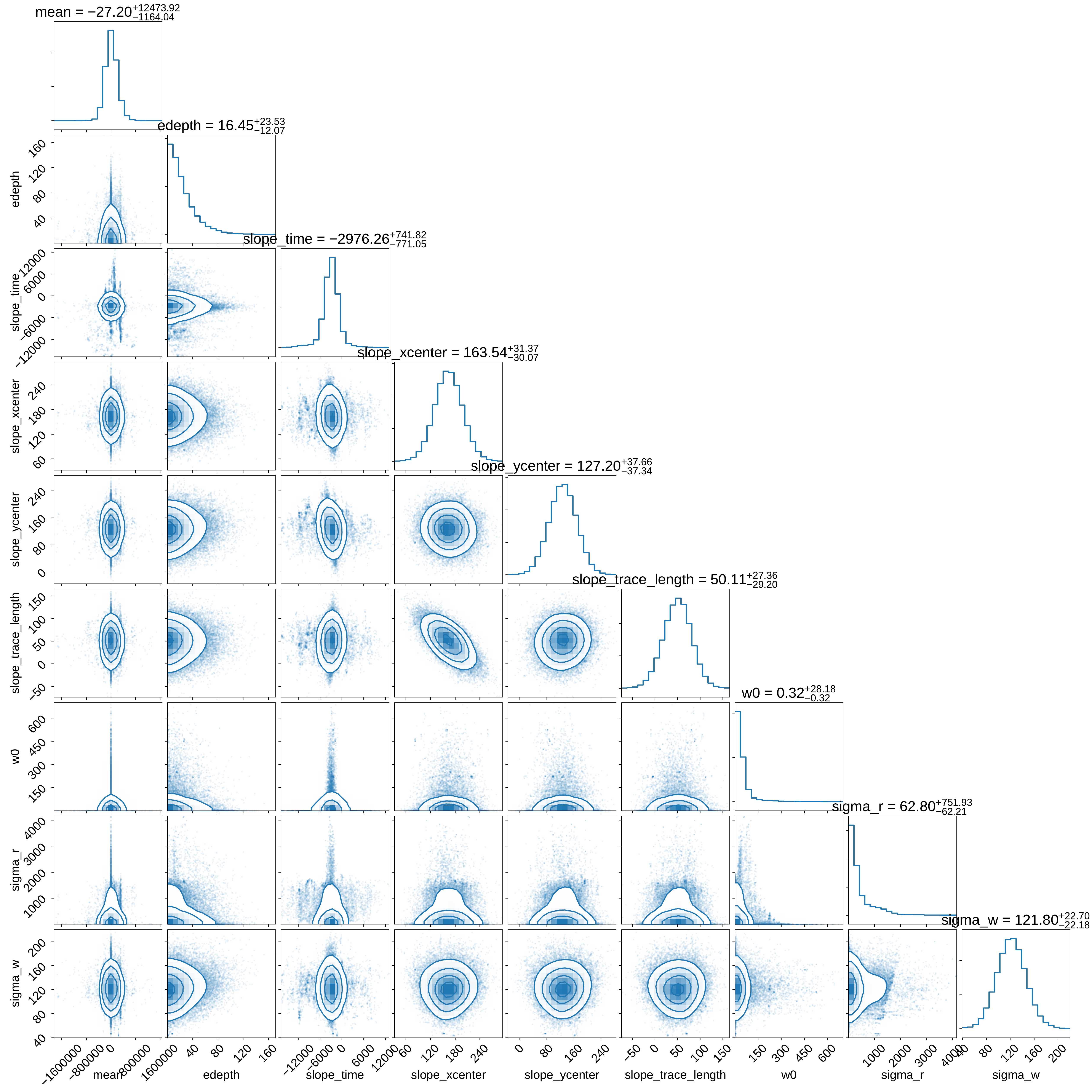}
\caption{Correlation (`corner') plot for our re-analysis using Gaussian Processes (GPs).
From personal communications with the \texttt{exoplanet} development team\footnote{https://github.com/exoplanet-dev/exoplanet/issues/76}, we used their built-in GP methods (i.e.~\texttt{celerite}) to create a GP that would probe the auto-correlation time scales and amplitudes for non-linear correlated noise in our residuals.
We simultaneously sampled the linear slopes for Time, X-Center, Y-Center, Trace-Lengths (same as our \textit{best} model above); along with the GP kernel hyper-parameters: $\sigma_r$, $\sigma_w$, $\omega_o$.
The result is two-fold: (1) the posterior for $\sigma_r$ is significantly skewed, with a median of 63$^{+751}_{-62}$~ppm and (2) \texttt{AICc}/\texttt{BIC} increased by $\Delta AICc \sim \Delta BIC > 10$.
Both results reject the hypothesis that power-law noise is measurable within our residuals.
}
\label{fig:celerite}
\end{figure}

We simultaneously fit the Time, X-Centers, Y-Centers, Trace-Lengths, and Offset linear models with the GP kernel hyper-parameters to constrain the Bayesian credible region over the new, non-linearly extended model. We provide a correlation plot for comparison in \autoref{fig:celerite}, which shows that the posterior over the power-law amplitude is significantly skewed and consistent with zero (the Null hypothesis).
\begin{equation*}
    \begin{array}{lccccccr}
         \sigma_w = 122 \pm 23 \text{ ppm} &&&&&& \sigma_r = 63^{+752}_{-62} \text{ ppm}
    \end{array}
\end{equation*}

Moreover, if we consider the three hyper-parameters of our kernel to be congruent with the 3-7 parameters of our parametric models (see \autoref{tbl:models}), then we also conclude that introducing the GP significantly increases the \texttt{AICc} and \texttt{BIC} by $\Delta AICc \sim \Delta BIC > 10$.
Altogether, we reject the hypothesis that auto-correlated noise exists in our residuals.
We further re-examined 6400 of our best non-GP included models by adding the same GP covariance matrix above.
We computed the MAP (not MCMC) for each of them, and also found that the use of GPs is not warranted, within the uncertainties, for our residuals.

\clearpage
\bibliographystyle{aasjournal}
\bibliography{references}
\end{document}